\documentclass[journal]{IEEEtran}
\usepackage{mathrsfs}
\usepackage{amsthm}
\usepackage{amsfonts}
\usepackage{graphicx,cite,epsfig,amssymb,amsmath}
\usepackage{color,xcolor}
\usepackage{pifont}
\usepackage{stmaryrd}
\usepackage{setspace}
\usepackage{subfigure}
\usepackage{cite}
\usepackage{amsmath}
\usepackage{array}\usepackage{float}\usepackage{multirow}
\usepackage{algorithmic,algorithm}
\usepackage{bm}
\usepackage{stfloats}

\floatstyle{ruled}
\newfloat{algorithm}{tbp}{loa}
\providecommand{\algorithmname}{Algorithm}
\floatname{algorithm}{\protect\algorithmname}

\newtheorem{definition}{Definition}
\newtheorem{assumption}{Assumption}

\ifCLASSINFOpdf

\else

\fi

\begin{document}

\title{URLLC and eMBB Coexistence in MIMO Non-orthogonal Multiple Access Systems }
\author{Qimei Chen, Jiajia Wang, and Hao Jiang
            \thanks{Q. Chen, J. Wang, and H. Jiang are with the School of Electronic Information, Wuhan University, Wuhan, China. e-mail: chenqimei@whu.edu.cn, wangjj@whu.edu.cn, jh@whu.edu.cn.}
            }

\maketitle

\begin{abstract}
Enhanced mobile broadband (eMBB) and ultra-reliable and low-latency communications (URLLC) are two major expected services in the fifth-generation mobile communication systems (5G).
Specifically, eMBB applications support extremely high data rate communications, while URLLC services aim to provide stringent latency with high reliability communications. Due to their differentiated quality-of-service (QoS) requirements, the spectrum sharing between URLLC and eMBB services becomes a challenging scheduling issue.
In this paper, we aim to investigate the URLLC and eMBB co-scheduling/coexistence problem under a puncturing technique in multiple-input multiple-output (MIMO) non-orthogonal multiple access (NOMA) systems.
The objective function is formulated to maximize the data rate of eMBB users while satisfying the latency requirements of URLLC users through joint user selection and power allocation scheduling.
To solve this problem, we first introduce an eMBB user clustering mechanism to balance the system performance
and computational complexity. Thereafter, we decompose the original problem into two subproblems, namely the scheduling problem of user selection and power allocation. We introduce a Gale-Shapley (GS) theory to solve with the user selection problem, and a successive convex approximation (SCA) and a difference of convex (D.C.) programming to deal with the power allocation problem. Finally, an iterative algorithm is utilized to find the global solution with low computational complexity. Numerical results show the effectiveness of the proposed algorithms, and also verify the proposed approach outperforms other baseline methods.
\end{abstract}

\maketitle

%\vspace*{-1cm}

\section{Introduction}
The fifth-generation (5G) wireless communication classifies the various service requirements into three types, namely enhanced Mobile Broad Band (eMBB), massive Machine Type Communications (mMTC), and Ultra Reliable Low Latency Communications (URLLC) \cite{5G-tutorial,5G-newradio}. Precisely, eMBB users require extremely high data rates, URLLC users concern high reliability and remarkably low latency, and mMTC users focus on a bulk of connections. Generally, the eMBB users would occupy the vast majority of wireless resources, whereas the URLLC users occur spontaneously. The most toilless way to deal with this matter is to reserve dedicated resources for URLLC users. Intuitively, this approach may lead to the wireless resources under-utilized.
Therefore, it is important to investigate the URLLC and eMBB coexistence issue on the same  wireless resources.

3GPP has introduced a superposition/puncturing and the short transmission time interval (short-TTI) technologies \cite{biaozhun,123}.
The main idea of these two technologies is to divide each transmission slot ($1$ ms) into several mini slots (0.125 ms), and then allow arrival URLLC users occupy the ongoing eMBB users. The transmission power of these interrupted eMBB users at specific mini slots would be set as zero as long as any URLLC user arrives.
Meanwhile, academia and industries have also paid much attention on this issue. Specifically, the authors in \cite{jianhua-tang-urllc-embb-slice} propose a slicing architecture for orthogonal URLLC and eMBB
coexistence, where URLLC slices remain unchanged before each transmission period.
An optimal joint scheduling problem for URLLC and eMBB traffic is proposed in \cite{urllc-embb-anand}, where linear, convex, and threshold based rate loss models are introduced.
The authors in \cite{chenlaoshi} propose a URLLC and eMBB coexistence mechanism on a cellular vehicle-to-everything (C-V2X) framework.
In \cite{spatial-embb-urllc}, the authors propose a null space based spatial preemptive scheduling for URLLC and eMBB traffic in a dense multi user 5G network. Furthermore, the authors in \cite{DRL-embb-urllc} model the URLLC and eMBB coexistence problem as a deep reinforcement learning issue.
Nevertheless, it is still an open issue to efficiently coexist URLLC and eMBB users.

On the other aspect, non-orthogonal multiple access (NOMA) has recently emerged as a promising access technology, due to its potential in achieving massive connections, user fairness, and high spectral efficiency \cite{BoyaDi-NOMA}.
Several works have already leveraged the NOMA technology into the URLLC and eMBB coexistence system. In \cite{R1-NOMA-SHENGAO}, the authors propose a NOMA access based multi-cell C-RAN architecture, where performance tradeoffs between URLLC and eMBB users have been investigated. Moreover, the authors in \cite{R2-NOMA-SHENGAO} consider the coexistence between URLLC and eMBB users in an analog C-RAN architecture, where uplink NOMA transmission have been studied. Furthermore, the authors in \cite{noma-gongcun} propose a NOMA based Fog-Radio architecture for URLLC and eMBB coexistence, which can schedule eMBB traffic centrally at the cloud and deal with URLLC traffic at the edge nodes. However, the above URLLC and eMBB coexistence systems have not considered the multiple-input-multiple-output (MIMO)-NOMA access, which is a more attractive technology for further spectrum efficiency enhancement through increasing the degree of spatial freedom.
%Apart from NOMA, multiple-input-multiple-output (MIMO) is another hopeful technology for further spectrum efficiency enhancement through increasing the degree of spatial freedom \cite{MIMO1}.
%Spontaneously, MIMO-NOMA becomes an attractive technology.
However, the development of MIMO-NOMA technology also confronts several challenges, such as the beamforming management and power allocation. In \cite{MIMO-NOMA1}, the authors propose a power allocation scheme to maximize the data rate of MIMO-NOMA system. In addition, the authors in \cite{MIMO-NOMA2} compare the performance of MIMO-NOMA with MIMO orthogonal multiple access (MIMO-OMA), and present the impressive gain achieved by MIMO-NOMA.
Moreover, the authors in \cite{Zhiguo-Ding-application-NOMA,fairness-of-user-cluster-NOMA} propose their MIMO-NOMA architectures based on a clustering technology. To the best of our knowledge, there has no work leveraging MIMO-NOMA into the URLLC and eMBB coexistence issue yet.
Since the MIMO-NOMA based system has a relatively high spatial freedom degrees and can connect more users, it provides more access opportunities for the arrival URLLC users as well as minimizes the performance loss of the connected eMBB users. Specifically, each arrival URLLC users can select a more suitable access position from the large number of connected eMBB users. Meanwhile, the connected eMBB users can also provide more eMBB users with worse performance to be selected.

%Since the MIMO-NOMA based system has a relatively high spatial freedom degrees, it provides more access opportunities for the arrival URLLC users as well as minimizes the performance loss to the connected eMBB users.

Based on the above analyses, in this paper, we introduce an efficient URLLC and eMBB coexistence mechanism in a downlink MIMO-NOMA system. Firstly, we introduce a dynamic user clustering mechanism before each transmission period. A maximal-minimum clustering algorithm is investigated to ensure the tradeoff between system performance and computational complexity.
Thereafter, we allow the arrived URLLC users to puncture the following mini slot immediately by interrupting eMBB transmissions in related clusters.
The URLLC and eMBB coexistence issue is formulated as a joint user selection and power allocation scheduling problem, which is a NP-hard mixed integer nonlinear programming (MINLP).
To deal with the optimization problem, we decompose it into two subproblems, namely punctured eMBB user selection and power allocation. We formulate the punctured user selection subproblem as a Gale-Shapley (GS) matching process, which provides an adaptive and low-complexity framework. Alternatively, the power allocation problem is simplified with a successive convex approximation (SCA) and a difference of convex (D.C.) programming.
The suboptimal solution of the whole problem is found through iteratively solving the above two subproblems. %Finally, we present the closed form average transmission data rate expressions of both URLLC and eMBB users, which highlights some insights.
%To the best of our knowledge, this is the first work considering URLLC and eMBB coexistence in MIMO-NOMA systems.

%\textcolor{red}{According to \cite{Zhiguo-Ding-application-NOMA,fairness-of-user-cluster-NOMA}, XXXXX
%However, it has been a concern that the decoding complexity and the implementation complexity of successive interference cancellation (SIC)
%increase with the growing number of users in MIMO-NOMA system. Furthermore, the error propagation has been an issue in SIC decoding.
%According to \cite{BoyaDi-NOMA}, SIC performed at the user receiver may cause considerable
%complexity ${\rm \mathcal {O}}\left ({ { {{K}} }^{3} }\right )$, where ${K}$ is the number of users in NOMA system.
%For example, considering a system with $3M$ users, we divide these users into $M$ clusters, each of which contains three users. Then,
%the number of users of messages decoded by SIC is reduced to 3.
%Thus the decoding complexity of SIC goes from ${\rm \mathcal {O}}\left ({ \left( {{ 3M}}\right) ^{3} }\right )$ to ${\rm \mathcal {O}}\left ({ { {{ 3}} }^{3} }\right )$.
%So computational complexity in clustering refers to the decoding complexity and the implementation complexity of SIC.}
%the computational complexity of a MIMO-NOMA system would incredibly increase with the number of users. %Therefore, user clustering technologies are indispensable in MINO-NOMA systems.
%In this way, the authors in \cite{access-NOMA} propose a user clustering technology to strike a balance between system performance and computational complexity.

The main contributions of this paper are summarized as follows.
\begin{enumerate}
  \item We propose an efficient URLLC and eMBB coexistence mechanism under a puncturing technology in MIMO-NOMA systems, which provides more flexibility for the access of arrival URLLC users and is more friendly to the existing eMBB users. In detail, any arrival URLLC users would replace the transmission of parts eMBB users in different spatial directions for the following mini slot.

  \item We formulate the objective function to maximize the eMBB data rates while ensuring the URLLC latency requirements. A joint user selection and power allocation scheme is proposed to improve the spectrum efficiency.

  \item An eMBB user clustering mechanism in MIMO-NOMA systems is introduced to not only ensure a stable eMBB user structure before transmission, but also strike a balance between system performance and computational complexity.

  \item The MINLP optimization problem is decomposed into two subproblems: punctured eMBB user selection and power allocation. We adopt a GS matching algorithm to solve the user selection problem, and a SCA-DC algorithm to deal with the power allocation problem.
      %Closed form expressions for the transmission data rates of URLLC and eMBB users are also presented.
\end{enumerate}

The rest of this paper is organized as follows. Section~\ref{A} presents the system model. Section~\ref{problem} is the problem formulation.
Section~\ref{B} and Section~\ref{C} propose the proposed user clustering and joint scheduling problems, respectively. Simulation results are presented in Section~\ref{D}. Finally, a conclusion is summarized in Section~\ref{E}.

\textit{Notations}:  The superscripts $\boldsymbol{A}^T$ and $ \boldsymbol{A}^H$ denote its transpose and conjugate transpose, respectively. Bold lowercase and uppercase letters represent column vectors and matrices, respectively.
$\mathbb{C}^{N\times M}$ denotes the set of all $N\times M$ matrices with complex entries.
$\mathbb {R}^{N \times M}$ denotes the set of all $N\times M$ matrices with real entries.
$\left |\cdot \right |$ denotes the absolute values of a complex scalar or the cardinality of a set. $\| \cdot \|$ represents the vector norm.

\section{System Model}\label{A}
\begin{figure}
  \centering
  \includegraphics[width=0.45\textwidth]{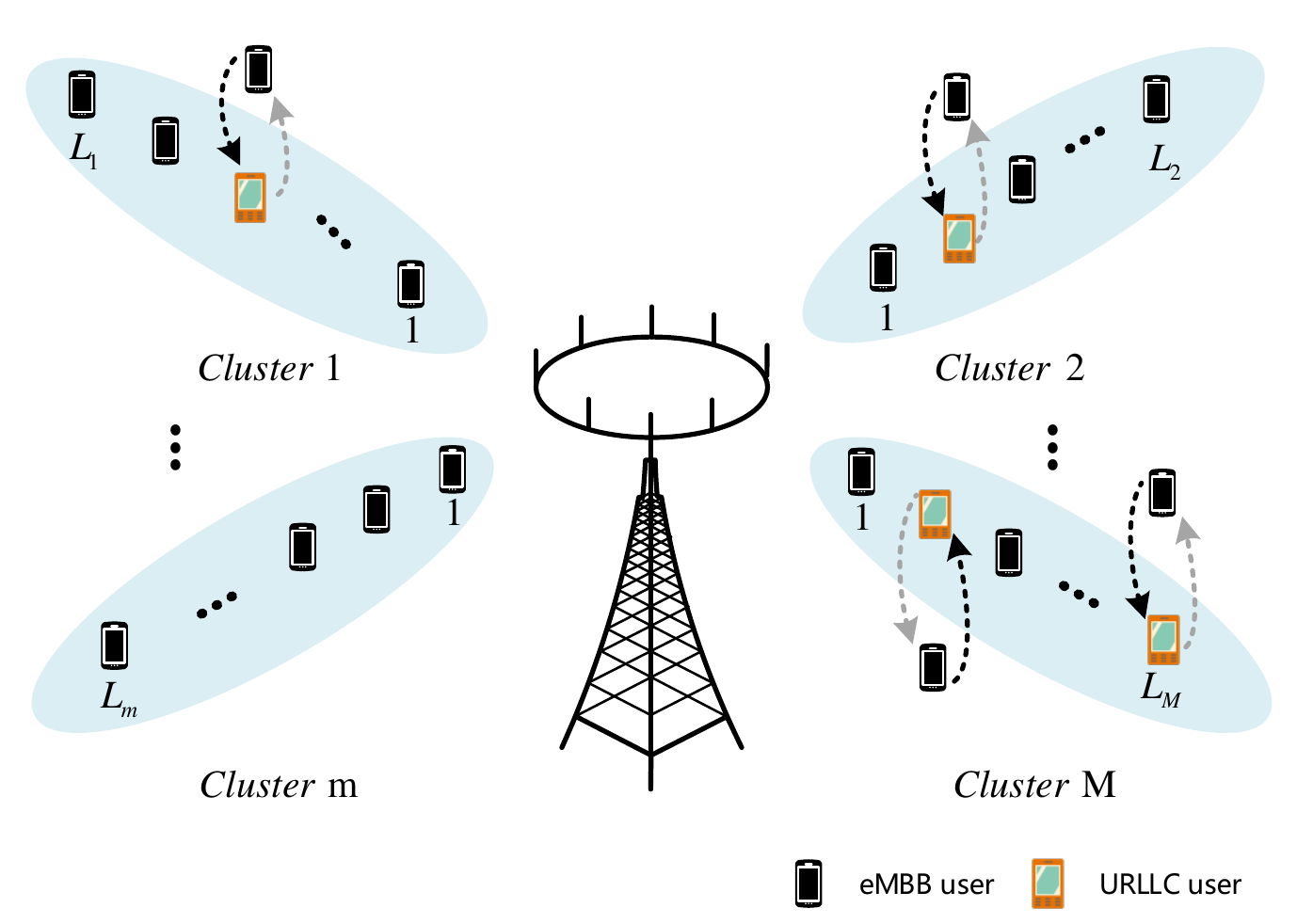}\\
  \caption{An illustration of the URLLC and eMBB coexistence in MIMO-NOMA systems.}\label{fig:1}
\end{figure}

In this work, we consider a URLLC and eMBB coexistence scenario under a downlink MIMO-NOMA based network.
There are $K$ eMBB devices and several URLLC devices within the coverage of a multi-antenna base station (BS), as shown in Fig.\ref{fig:1}.
The BS equips with $B$ antennas, and each eMBB or URLLC user equips with $N$ antennas.
Each time slot (transmission period of the framework) is divided into $Q$ equivalent mini slots, denoted as $\mathcal {Q}\triangleq \{1,\cdots,Q\}$, where $i \in \mathcal {Q}$ represents the $i$-th mini slot.

Under the MIMO-NOMA system, $K$ eMBB users are grouped into $M$ clusters before each transmission period, denoted as ${\mathcal{M}}=\left \{{ 1,2,\ldots,M }\right \}$, where the number of clusters equals to the number of BS antennas, i.e., $M=B$.
%and $m \in {\mathcal{M}}$ represents the $m$-th cluster in MIMO-NOMA system.
Furthermore, we denote ${\mathcal{L}} = \left \{{ {\mathcal{L}_{1}},{\mathcal{L}_{2}},\ldots,{\mathcal{L}_{M}} }\right \}$,
where ${\mathcal{L}_{m}} \in {\mathcal{L}}$ represents the user set in the $m$-th cluster.
Cluster ${\mathcal{L}_{m}}$ contains ${L_{m}=\left |{ \mathcal{L}_{m}}\right |}$ users.
Here, we ensure each cluster contains at least two users, i.e., ${\left |{ \mathcal{L}_{m}}\right |}\geq 2$.

\subsection{MIMO-NOMA Models}
Based on \cite{MIMO-NOMA-gainian}, the considered multi-user MIMO-NOMA scenario can be decomposed into multiple separate single-antenna NOMA channels, to which conventional NOMA technology can be applied directly by reasonable power allocation and user clustering.
The downlink MIMO-NOMA transmission is implemented by superimposing the signals destined for end users at the same cluster.
As a result, the transmit signal vector $\tilde {\mathbf {s}}^{(i)} \in \mathbb{C}^{M\times 1}$ can be written as
\begin{equation}\label{eq_s}
\tilde {\mathbf {s}}^{(i)} =\left [{\! {\begin{array}{*{20}{c}} {\sqrt {{\alpha _{1,1}^{(i)}} }{s_{1,1}^{(i)}} + \cdots + {\sqrt {{\alpha _{1,{L_{1}}}^{(i)}}} {s_{1,{L_{1}}}^{(i)}}}} \\ \vdots \\ {\sqrt {{\alpha _{M,1}^{(i)}}} {s_{M,1}^{(i)}} + \cdots + {\sqrt {{\alpha _{M,{L_{M}}}^{(i)}}}}} {s_{M,{L_{M}}}^{(i)}} \\ \end{array}} \!}\right ] \buildrel \Delta \over = \left [{\! {\begin{array}{*{20}{c}} {{{\tilde {\mathbf{s}}}_{1}^{(i)}}}\\ \vdots \\ {\tilde {\mathbf { {s}}}_{M}^{(i)} } \end{array}} \!}\right ]\!,\qquad \end{equation}
where $s_{m,k}^{(i)}$ and $\alpha _{m,k}^{(i)}$ are the transmitted information and the NOMA power allocation coefficient of the $k$-th user in the $m$-th cluster at mini slot $i$, respectively.

The BS precodes the signal vector with an identity precoding matrix $\textbf {P}^{(i)} \in \mathbb{C}^{M\times M}$\cite{Zhiguo-Ding-application-NOMA}.
Hence, the received signal of the $k$-th user over the $m$-th cluster in mini slot $i$ can be expressed as
\begin{equation}
{\textbf {y}_{m,k}^{(i)}} = {\textbf {H}_{m,k}^{(i)}}\textbf {P}^{(i)} \tilde {\mathbf {s}}^{(i)} + {\textbf {n}_{m,k}^{(i)}},~ \forall m,k\in {\mathcal{L}_{m}},i,
\end{equation}
where
%$\textbf {P}^{(i)} \in \mathbb{C}^{M\times M}$ is a identity precoding matrix at \textcolor{red}{transmitting terminals??? the BS?},
$\textbf {H}_{m,k}^{(i)} \in \mathbb{C}^{N\times M}$ is the channel gain matrix from the BS to the $k$-th user in the $m$-th cluster, and ${\textbf {n}_{m,k}^{(i)}}$ is additive white Gaussian noise (AWGN) vector. For simplicity, the channel gains between the BS and users are assumed to be independent and identically complex Gaussian distributed\footnote{Note that there has no relationship between the BS/user's channel gain distribution and the channel spatial correlation.}. Moreover, the channel state is invariant during each mini slot and time-varying across different mini slots.

To further improve the spectral efficiency, a detection vector $\textbf v_{m,k}^{(i)} \in \mathbb{C}^{N\times 1}$ has been applied
by each user.
%\textcolor{red}{receiving terminals/the users???}.
Therefore, the detection results of the $k$-th user in the $m$-th cluster can be formulated as
\begin{multline}\label{vy}
{\textbf {v}_{m,k}^{(i)~H}}{{\textbf {y}_{m,k}^{(i)}}} = {\textbf {v}_{m,k}^{(i)~H}}{\textbf {H}_{m,k}^{(i)}}\textbf {P}^{(i)}\tilde {\mathbf {s}}^{(i)} + {\textbf {v}_{m,k}^{(i)~H}}{\textbf {n}_{m,k}},\\
\forall m,k\in {\mathcal{L}_{m}},i.
\end{multline}

Denote ${{\mathbf{p}}_{m}^{(i)}}$ as the $m$-th column of $\textbf {P}^{(i)}$, \eqref{vy} can be rewritten as
\begin{equation}
\begin{split}
  {\textbf {v}_{m,k}^{(i)~H}}{{\mathbf{y}}_{m,k}^{(i)}}
= \sum \limits _{f = 1,f \ne m}^{M} {{\textbf {v}_{m,k}^{(i)~H}}
{{\mathbf{H}}_{m,k}^{(i)}}
{{\mathbf{p}}_{f}^{(i)}}
{{\tilde {\mathbf{s}}}_{f}^{(i)}}} +
{\textbf {v}_{m,k}^{(i)~H}}{\textbf {n}_{m,k}^{(i)}}
+\\
{\textbf {v}_{m,k}^{(i)~H}}{{\mathbf{H}}_{m,k}^{(i)}}{{\mathbf{p}}_{m}^{(i)}}\bigg ({ {\sqrt {{\alpha _{m,1}^{(i)}}} {s_{m,1}^{(i)}} + \cdots    } }
{ { + \sqrt {{\alpha _{m,{L_{m}}}^{(i)}}} {s_{m,{L_{m}}}^{(i)}}} }\bigg ),\\
~ \forall m,k\in {\mathcal{L}_{m}},i.
\end{split}
\end{equation}

To remove the inter-cluster interference, the precoding and detection matrices should satisfy the constraints of
\begin{multline}
{ \label{10} {\mathbf{v}_{{l,k}}^{(i)~H}}}\underbrace{\left[{\mathbf{h}_{1,lk}^{(i)}\cdots \mathbf{h}_{l-1,lk}^{(i)}\;\mathbf{h}_{l+1,lk}^{(i)}\cdots\mathbf{h}_{M,lk}^{(i)}}\right]}_{\tilde{\mathbf{H}}_{l,k}^{(i)}}=0,
\\
\forall l \in \mathcal{M}, k\in \mathcal{L}_{\ell},i,
\end{multline}
where ${{\mathbf{h}}_{j,lk}^{(i)}}$ refers to the $j$-th column of ${\textbf {H}_{l,k}^{(i)}}$, $\tilde{\mathbf{H}}_{l,k}^{(i)} \in \mathbb{C}^{N\times (M-1)}$ is a submatrix of $\mathbf{H}_{l,k}^{(i)}$ by removing the $l$-th column.

%\begin{equation}\label{no_interference}
%\mathbf{v}_{l,k}^{(i)~H}\mathbf{h}_{j,lk}^{(i)}=0,~ \forall l \in \mathcal{M}, k\in \mathcal{L}_{\ell}, i, j\neq l.
%\end{equation}
%Here, ${{\mathbf{h}}_{j,lk}^{(i)}}$ indexes to the $j$-th column of ${\textbf {H}_{l,k}^{(i)}}$.  Constraints \eqref{no_interference} can be further rearranged as
%\begin{multline}
%{ \label{10} {\mathbf{v}_{{l,k}}^{(i)~H}}}\underbrace{\left[{\mathbf{h}_{1,lk}^{(i)}\cdots \mathbf{h}_{l-1,lk}^{(i)}\;\mathbf{h}_{l+1,lk}^{(i)}\cdots\mathbf{h}_{M,lk}^{(i)}}\right]}_{\tilde{\mathbf{H}}_{l,k}^{(i)}}=0,
%\\
%\forall l \in \mathcal{M}, k\in \mathcal{L}_{\ell},i,
%\end{multline}
%where $\tilde{\mathbf{H}}_{l,k}^{(i)} \in \mathbb{C}^{N\times (M-1)}$ is a submatrix of $\mathbf{H}_{l,k}^{(i)}$ by removing the $l$-th column.}

Consequently,
$\mathbf{v}_{l,k}^{(i)}$ can be obtained from the null space of $\tilde{\mathbf{H}}_{l,k}^{(i)}$, i.e., $\mathbf{v}_{l,k}^{(i)}=\mathbf{U}_{l,k}^{(i)}\mathbf{z}_{l,k}^{(i)}$, where $\mathbf{U}_{l,k}^{(i)}$ contains all the left singular vectors of $\tilde{\mathbf{H}}_{l,k}^{(i)}$ corresponding to zero singular values, and $\mathbf{z}_{l,k}^{(i)} \in \mathbb{C}^{(N-M+1)\times 1}$ is a normalized vector.
According to \cite{Zhiguo-Ding-application-NOMA} and \cite{fairness-of-user-cluster-NOMA},  $\mathbf{z}_{l,k}^{(i)}$ can be obtained by using the maximal ratio combining (MRC) approach, which is given by
\begin{equation}\label{9}
\mathbf{z}_{l,k}^{(i)} = \frac{\mathbf{U}_{l,k}^{(i)~\!\!H} \mathbf{h}_{l,lk}^{(i)}}{\| \mathbf{U}_{l,k}^{(i)~\!\!H} \mathbf{h}_{l,lk}^{(i)}\|},~\forall l \in \mathcal{M}, k\in \mathcal{L}_{\ell},i.
\end{equation}

To reduce system overheads, we allow the BS transmits signals without processing them that can avoid the channel state information (CSI) feedback. However, to implement the successive interference cancellation (SIC) effectively, the BS needs to obtain the information of the channel gain order.
We utilize the coefficients of ${\left |{ {{\textbf {v}_{m,k}^{(i)~H}}{{\mathbf{H}}_{m,k}^{(i)}}{{\mathbf{p}}_{m}^{(i)}}} }\right |^{2}}$
to describe the channel conditions, which can be generally ordered as
\begin{equation} \label{channel_order}
{\left |{ {{\textbf {v}_{m,1}^{(i)~H}}{{\mathbf{H}}_{m,1}^{(i)}}{{\mathbf{p}}_{m}^{(i)}}} }\right |^{2}} \ge \cdots \ge {\left |{ {{\mathbf{v}}_{m,{L_{m}}}^{(i)~H}{{\mathbf{H}}_{m,{L_{m}}}^{(i)}}{{\mathbf{p}}_{m}^{(i)}}} }\right |^{2}},~ \forall m,i.
\end{equation}
Note that, we would sort the effective channel gain at the beginning of each mini slot.

According to the SIC, user $k$ in the $m$-th cluster needs to decode the messages of the users with poorer channel conditions in the same cluster $s_{m,j}^{(i)}, ~k\leq j \le {{ L_{m}}}$ before detecting its own information $s_{m,k}^{(i)}$.
The signal $s_{m,j}^{(i)}$ can be detected at the $k$-th user with the following SINR as
\begin{multline}\label{chushiSINR}
SINR_{m,k,j}^{(i)} \!=\! \frac {{{{\left |{ {{\mathbf{v}}_{m,k}^{(i)~H}{{\mathbf{h}}_{m,mk}^{(i)}}} }\right |}^{2}}{\alpha _{m,j}^{(i)}}}}{{\sum \nolimits _{l = 1}^{j - 1} {{{\left |{ {{\mathbf{v}}_{m,k}^{(i)~H}{{\mathbf{h}}_{m,mk}^{(i)}}} }\right |}^{2}}{\alpha _{m,l}^{(i)}} + {{ \|{ {{{\mathbf{v}}_{m,k}^{(i)}}} }\|}^{2}}\frac {1}{\rho }}}},\\
\forall m,  k\in {\mathcal{L}_{m}} ,
 j \in \left \{{ k,k+1,\ldots,{L_{m}} }\right \},i,
\end{multline}
where $\rho$ denotes the transmit signal-to-noise-ratio (SNR).

Therefore, the effective SINR for user $k$ in the $m$-cluster at mini slot $i$ can be expressed as \cite{BoyaDi-NOMA}
\begin{multline} \label{sinr}
{SINR_{m,k}^{(i)}}=
\frac {{{{\left |{ {{\mathbf{v}}_{m,k}^{(i)~H}{{\mathbf{h}}_{m,mk}^{(i)}}} }\right |}^{2}}{\alpha _{m,k}^{(i)}}}}{{\sum \nolimits _{l = 1}^{k - 1} {{{\left |{ {{\mathbf{v}}_{m,k}^{(i)~H}{{\mathbf{h}}_{m,mk}^{(i)}}} }\right |}^{2}}{\alpha _{m,l}^{(i)}} + {{\|{ {{{\mathbf{v}}_{m,k}^{(i)}}} }\|}^{2}}\frac {1}{\rho }}}},\\
\forall m,k\in {\mathcal{L}_{m}},i.
\end{multline}

\newcounter{mytempeqncnt1}
\subsection{URLLC and eMBB Transmission Models}
To ensure the strict latency requirement, any arrival URLLC user is allowed to puncture the next mini slot immediately by interrupting an eMBB user's transmission in a specific cluster. Meanwhile, the BS allocates zero transmission power to the punctured eMBB user \cite{urllc-embb-anand}. After the puncturing, the interrupted eMBB user would reoccupy the mini slot again.
Denote $\mathcal{S}^{(i)}_{~\!u}$ as the set of arrival URLLC users in mini slot $i$, which would interrupt the transmission of $\left |{  \mathcal{S}^{(i)}_{~\!u}}\right |$ eMBB users.
Let $\mathcal{S}^{(i)}_{~\!u,m}$ be the set of URLLC users puncturing in cluster $m$ at mini slot $i$.
Then, we have
${\mathcal{S}^{(i)}_{~\!u}}=\mathcal{S}^{(i)}_{~\!u,1}\cup \mathcal{S}^{(i)}_{~\!u,2} \cup \cdots \cup \mathcal{S}^{(i)}_{~\!u,M}$. Denote $\mathcal{S}^{(i)}_{~\!e,m}$
as the set of remaining eMBB users for cluster $m$ at mini slot $i$, and $\mathcal{S}^{(i)}_{~\!e}$ as the set of all the remaining eMBB users at mini slot $i$, where
$\left |{  \mathcal{S}^{(i)}_{~\!e}}\right |=K-{\left |{  \mathcal{S}^{(i)}_{~\!u}}\right |}$ and  $\left |{  \mathcal{S}^{(i)}_{~\!e,m}}\right |={L_{m}}
-{\left |{  \mathcal{S}^{(i)}_{~\!u,m}}\right |}$.
Let $x _{m,k,n}^{(i)}$ be a binary variable to indicate whether eMBB user $k$ in $m$-th cluster is punctured by the URLLC user $n$. $x _{m,k,n}^{(i)}=1$ represents eMBB user $k$ in the $m$-th cluster is punctured by URLLC user $n$, and $x _{m,k,n}^{(i)}=0$ otherwise.

We further define $\beta _{m,n}^{(i)}$ and $\textbf {w}_{m,n}^{(i)}$ as the power allocation coefficient and detection vector of the $n$-th URLLC user in the $m$-th cluster at mini slot $i$, respectively. After  puncturing, the effective SINR for URLLC user $n \in \mathcal{S}^{(i)}_{~\!u,m}$ in the $m$-cluster at mini slot $i$ can be rewritten as
\begin{equation}\label{sinr_urllc}
 SINR_{u,m,n}^{(i)}
= \frac {{{{\left |{ {{\mathbf{w}}_{m,n}^{(i)~H}{{\mathbf{h}}_{m,mn}^{(i)}}} }\right |}^{2}}{\beta _{m,n}^{(i)}}}}
{I_{1}^{u}+I_{2}^{u}+\! {{\|{ {{{\mathbf{w}}_{m,n}^{(i)}}} }\|}^{2}}\frac {1}{\rho }}, ~\forall m, n \in \mathcal{S}^{(i)}_{~\!u,m}, i,
\end{equation}
where
\begin{equation}
I_{1}^{u}=\sum \limits_{l = 1}^{j - 1} (1-\!\sum \limits _{n^{'} \in \mathcal{S}^{(i)}_{~\!\!u,m}}\!x _{m,l,n^{'}}^{(i)}){{{\left |{ {{\mathbf{w}}_{m,n}^{(i)~H}{{\mathbf{h}}_{m,mn}^{(i)}}} }\right |}^{2}}{\alpha _{m,l}^{(i)}}},
\end{equation}
and
\begin{equation}
I_{2}^{u}=\sum \limits_{l = 1}^{j - 1}
\sum \limits _{n^{'} \in \mathcal{S}^{(i)}_{~\!\!u,m}}\!{x _{m,l,n^{'}}^{(i)}}{{{\left |{ {{\mathbf{w}}_{m,n}^{(i)~H}{{\mathbf{h}}_{m,mn}^{(i)}}} }\right |}^{2}}{\beta _{m,l}^{(i)}}}.
\end{equation}

Since URLLC packets are typically quite short, the achievable rate cannot be accurately captured by Shannon's capacity. Derived by \cite{5452208}, the data rate of URLLC user $n$ in $m$-th cluster at mini slot $i$ falls into the finite blocklength channel coding regime, which can be formulated as
\begin{equation}
\begin{split}\label{urllcrate}
r_{m,n}^{(i)}\leq {\log_{2}} (1+ {SINR}_{u,m,n}^{(i)})- \sqrt {\frac {C_{m,n}^{(i)}}{b_{m,n}^{(i)}}}Q^{-1}(\epsilon)\log e,\\
 \forall m, n \in \mathcal{S}^{(i)}_{~\!u,m}, i,
\end{split}
\end{equation}
where $Q^{-1}(\cdot)$ is the inverse of the Gaussian Q-function, $\epsilon >0$ is the transmission error probability, $b_{m,n}^{(i)}$ is the length of codeword block in symbols, and ${C_{m,n}^{(i)}}$ is the channel dispersion with
\begin{equation}\label{channel_dis}
C_{m, n}^{(i)}=1-\frac {1}{(1+{SINR}_{u,m,n}^{(i)})^{2}},~ \forall m, n\in \mathcal{S}^{(i)}_{~\!u,m}, i.
\end{equation}

According to \cite{code1,code2}, the upper bound of \eqref{urllcrate} can be achieved through a suitable encoding/decoding couple. Therefore, we set the upper bound for $r_{u,m,n}^{(i)}$ in the following paper.

To satisfy the latency requirements, URLLC users should satisfy \cite{jianhua-tang-urllc-embb-slice}
\begin{equation}
\frac {F_{m,n}^{(i)}}{Br_{m,n}^{(i)}} \leq D_{m,n}^{(i), \max}, ~ \forall m,n \in \mathcal {S}^{(i)}_{~\!u,m},i,
\end{equation}
where $B$ is the bandwidth, $D_{m,n}^{(i), \max}$ and $F_{m,n}^{(i)}$ are the maximum tolerant delay and the packet length of URLLC user $n$ in the $m$-th cluster at mini slot $i$, respectively.

Similarly, the effective SINR of eMBB user $k \in \mathcal{S}^{(i)}_{~\!e,m}$ in the $m$-th cluster at mini slot $i$ can be expressed as
\begin{equation}\label{sinr_embb}
\begin{aligned}
 SINR_{e,m,k}^{(i)}
= \frac {{{{\left |{ {{\mathbf{v}}_{m,k}^{(i)~H}{{\mathbf{h}}_{m,mk}^{(i)}}} }\right |}^{2}}{\alpha _{m,k}^{(i)}}}}
{{{I_{1}^{e}+I_{2}^{e}+\|{ {{{\mathbf{v}}_{m,k}^{(i)}}} }\|}^{2}}\frac {1}{\rho }},~\forall m, k \in \mathcal{S}^{(i)}_{~\!e,m},i,
\end{aligned}
\end{equation}
where
\begin{equation}
\begin{aligned}
 I_{1}^{e}
=
\sum \limits_{l = 1}^{j - 1} (1-\!\sum \limits _{n \in \mathcal{S}^{(i)}_{~\!\!u,m}}\!x _{m,l,n}^{(i)}){{{\left |{ {{\mathbf{v}}_{m,k}^{(i)~H}{{\mathbf{h}}_{m,mk}^{(i)}}} }\right |}^{2}}{\alpha _{m,l}^{(i)}}},
\end{aligned}
\end{equation}
and
\begin{equation}
\begin{aligned}
I_{2}^{e}
=
\sum \limits_{l = 1}^{j - 1}
\sum \limits _{n \in \mathcal{S}^{(i)}_{~\!\!u,m}}\!{x _{m,l,n}^{(i)}}{{{\left |{ {{\mathbf{v}}_{m,k}^{(i)~H}{{\mathbf{h}}_{m,mk}^{(i)}}} }\right |}^{2}}{\beta _{m,l}^{(i)}}}.
\end{aligned}
\end{equation}

Thereafter, the data rate of eMBB user $k$ in the $m$-th cluster at mini slot $i$ can be formulated as
\begin{equation}
\begin{aligned}\label{embbrate}
{R_{m,k}^{(i)}}\!=\!{\log _{2}}\left({1\!+\!SINR_{e,m,k}^{(i)}}\right),~\forall m, k \in \mathcal{S}^{(i)}_{~\!e,m},i.
\end{aligned}
\end{equation}

To ensure the data rate requirements, eMBB users should satisfy
\begin{equation}
{R_{m,k}^{(i)}}\ge {R_{m,k}^{\min }},~\forall m, k \in \mathcal{S}^{(i)}_{~\!e,m},i,
\end{equation}
where ${R_{m,k}^{\min }}$ is the minimum throughput constraint of eMBB user $k$ in the $m$-th cluster.

\section{Problem Formulation}\label{problem}
In this section, we first formulate the user clustering problem. Thereafter, we introduce the joint user selection and power allocation problem.

\subsection{User Clustering}
To ensure the tradeoff between system performance and computational complexity, the user clustering problem is formulated to maximize the SINR of the worst user among all the eMBB users, which is a double maximize minimum problem of
\begin{align*}
&\mathcal {P}_{0}: \quad \max \limits _{\boldsymbol{\Lambda}}
\min \limits _{m}
{ {
{ {\max \limits _{{\boldsymbol{\alpha}}_{m}^{(i)}} \min \limits _{k,j} \left ({ {SINR_{m,k,j}^{(i)}} }\right )} }
} }
 , \tag{21}\label{19}
 \\&\quad \, \mathrm{subject}~\mathrm{to}
\\&\qquad \quad \log_2\left(1+SINR_{m,k,j}^{(i)}\right)\ge {R_{m,j}^{\min }},\\&
\qquad \quad\qquad
\forall m,  k\in {\mathcal{L}_{m}} ,
 j \in \left \{{ k,k+1,\ldots,{L_{m}} }\right \} , \tag{21a}\label{19a}
\\&\qquad\quad {\sum \limits _{k = 1}^{L_{m}} {{\alpha _{m,k}^{(i)}}} \leq \frac {L_{m}}{K},} ~ \forall m, \tag{21b}\label{19b}
       \\&\qquad\quad {\alpha _{m,k}^{(i)}} \ge 0,~ \forall m,k \in {\mathcal{L}_{m}}, \tag{21c}\label{19c}
\end{align*}
where $\boldsymbol{\Lambda}$ is defined as all possible clustering combination sets of all eMBB users, ${\boldsymbol{\alpha}}_{m}^{(i)}= \{\alpha _{m,1}^{(i)},...,\alpha _{m,L_{m}}^{(i)} \}, \forall m$ is the power allocation coefficient vector for the $m$-th cluster.
Constraints \eqref{19a} are the minimum data rate requirements. Constraints \eqref{19b}-\eqref{19c} are the total power limitations in different clusters.
%Note that we set the mini slot $i=1$, since the user clustering problem is related to the long time slot.

In detail, the inner maximize minimum problem
$\max \limits _{{\boldsymbol{\alpha}}_{m}^{(i)}} \min \limits _{k,j} \left ({ {SINR_{m,k,j}^{(i)}} }\right )$
is formulated to maximize the worst SINR in the $m$-th cluster at a given user clustering structure through power assignment.
Thereafter, the problem
$\min \limits _{m}
{ {
{ {\max \limits _{{\boldsymbol{\alpha}}_{m}^{(i)}} \min \limits _{k,j} \left ({ {SINR_{m,k,j}^{(i)}} }\right )} }
} }$
means to select the smallest SINR from all the $M$ clusters.
Finally, the outermost maximize problem $\max \limits _{\boldsymbol{\Lambda}}\min \limits _{m}
{ {
{ {\max \limits _{{\boldsymbol{\alpha}}_{m}^{(i)}} \min \limits _{k,j} \left ({ {SINR_{m,k,j}^{(i)}} }\right )} }
} }$ denotes to get the maximum value of the worst SINR among all clusters through traversing all user clustering combinations.

\subsection{Optimization Problem}
The goal of this work is to maximize the data rate of eMBB users during all the transmission period, while satisfying the QoS requirements of URLLC and eMBB users. Therefore, the optimal scheduling problem can be formulated as
\begin{align*}
&\mathcal {P}_{1}: ~ \max \limits _{\left \{{ x _{m,k,n}^{(i)}, {\boldsymbol {\alpha _{m}^{(i)}}},\boldsymbol {\beta _{m}^{(i)}}  }\right \}} \sum \limits _{i = 1}^{Q} \sum \limits _{m = 1}^{M} \sum \limits _{k = 1}^{{L_{m}}} \log_2\left ({ {1 +SINR_{e,m,k}^{(i)} } }\right ), \tag{22}\label{20}
\\& \mathrm{subject}~\mathrm{to}
\\& \qquad\quad \frac {F_{m,n}^{(i)}}{Br_{m,n}^{(i)}} \leq D_{m,n}^{(i), \max}, ~ \forall m, n \in \mathcal {S}^{(i)}_{~\!u,m}, i ,\tag{22a}\label{20a}
\\& \qquad\quad {R_{m,k}^{(i)}}\ge {R_{m,k}^{\min }},~ \forall m,  k \in \mathcal {S}^{(i)}_{~\!e,m},i, \tag{22b} \label{20b}
\\& \qquad\quad x _{m,k,n}^{(i)} \in \left \{{ {0,1} }\right \},~ \forall m,k \in \mathcal{L}_{m}, n \in \mathcal {S}^{(i)}_{~\!u,m},i , \tag{22c}\label{20c}
\\& \qquad\quad \sum \limits_{m \in {\mathcal{M}}}
\sum \limits _{k \in \mathcal{L}_{m}}x _{m,k,n}^{(i)} =1, ~ \forall n \in \mathcal{S}^{(i)}_{~\!u},  i, \tag{22d}\label{20d}
\\& \qquad\quad
\sum \limits _{n \in \mathcal{S}^{(i)}_{~\!u}}x _{m,k,n}^{(i)} \leq 1, ~ \forall m,k \in \mathcal{L}_{m},  i, \tag{22e}\label{20e}
\\& \qquad\quad {\alpha _{m,k}^{(i)}}>0, ~\forall m,k \in {S}_{~\!e,m}^{(i)},i, \tag{22f}\label{20f}
\\& \qquad\quad  {\beta _{m,n}^{(i)}}>0,~\forall m,n \in \mathcal {S}_{~\!u,m}^{(i)},i , \tag{22g}\label{20g}
\\& \qquad\quad \sum \limits _{k \in \mathcal{S}^{(i)}_{~\!\!e,m}}\!\!{\alpha _{m,k}^{(i)}}\!+\! \sum\limits _{n \in \mathcal{S}^{(i)}_{~\!\!u,m}}\!\!{\beta _{m,n}^{(i)}}
 \leq \frac{{{L_{m}}}}{K}, ~ \forall m ,i, \tag{22h}\label{20h}
\end{align*}
where ${\boldsymbol {\alpha}} _{m}^{(i)}=
\left \{{ {\alpha _{m,1}^{(i)}}, \cdots , {\alpha _{m,{\left |{  \mathcal{S}^{(i)}_{~\!\!e,m}}\right |}}^{(i)}} }\right \}$ and ${\boldsymbol {\beta }} _{m}^{(i)}=
\left \{{ {\beta _{m,1}^{(i)}}, \cdots , {\beta _{m,{\left |{  \mathcal{S}^{(i)}_{~\!\!u,m}}\right |}}^{(i)}} }\right \}$ are the power allocation coefficient vector of URLLC and eMBB users in the $m$-th cluster, respectively.
Constraints \eqref{20c}-\eqref{20e} ensure each URLLC user occupies one eMBB user and each eMBB user is punctured by at most one URLLC user.
Constraints \eqref{20f}-\eqref{20h} denote the transmit power limitation in each cluster.

Note that the computational complexity of the NOMA technology is based on the decoding and implementation complexities of the SIC, which increases with the number of users. Based on \cite{BoyaDi-NOMA}, the SIC complexity without user clustering is ${\rm \mathcal {O}}\left ({ { {{ K}} }^{3} }\right )$.
However, with user clustering, the computational complexity of SIC reduces to ${\rm \mathcal {O}}\left ({ \sum \nolimits _{m= 1}^{M}{\left(L_{m}\right)}^{3} }\right )$, where $K\!=\!\sum \nolimits _{m= 1}^{M}{L_{m}}$.

\section{User Clustering Mechanism}\label{B}
In this section, a user clustering mechanism would be proposed to solve $\mathcal {P}_{0}$.

Recalling \eqref{chushiSINR}, the objective user clustering function $\mathcal {P}_{0}$ is obviously a non-convex problem.
To solve this problem, we split the objective function \eqref{19} into two steps.

\begin{itemize}
  \item Step 1: Under a given eMBB user clustering combination set $\varphi \in \boldsymbol{\Lambda}$, $\mathcal {P}_{0}$ can be reformulated as
      \begin{align*}
      &\mathcal {P}_{2}: \quad
      \max \limits _{{\boldsymbol{\alpha}}_{m}^{(i)}} \min \limits _{k, j} \left ({ {\boldsymbol{SINR_{m}}}\left ({ {\boldsymbol{\alpha}}_{m}^{(i)} }\right ) }\right )
      , \tag{23}\label{22}
     \end{align*}
      subject to \eqref{19a}, \eqref{19b} and \eqref{19c}.

      Here, ${\boldsymbol{SINR_{m}}}$ denotes all possible values of $ SINR_{m,k,j}^{(i)}$, $\forall k,j$ in the $m$-th cluster under a given power allocation coefficient vector ${\boldsymbol{\alpha}}_{m}^{(i)}$.

      However, $\mathcal {P}_{2}$ is still non-convex, which leads us to find an equivalent conversions to make it tractable. Therefore, we define ${\mathbb {G}_\Theta }=\left \{{ {{\boldsymbol{SINR}}_{\mathbf{m}}\left ({{\boldsymbol{\alpha}}_{m}^{(i)}}\right ) \ge {\Theta}} }\right \}$ as the set of ${\boldsymbol{\alpha}}_{m}^{(i)}$ when the objective function is larger than $\Theta$, where $\Theta$ is a real number.
     Hence, $\Theta$ is equivalent to the minimum value of ${\boldsymbol{SINR_{m}}}$, i.e., $SINR_{\min}^{m}=\Theta$.
      Therefore, we can equally transform $\mathcal {P}_{2}$ as \cite{fairness-of-user-cluster-NOMA,max-min-reference}
      \begin{align*}
      &\mathcal {P}_{3}: \quad\quad\quad\quad \mathrm{Find} \quad \quad{{\boldsymbol{\alpha}}_{m}^{(i)}} ,\tag{24}\\&\quad\, \mathrm{subject}~\mathrm{to}
       \\&\qquad\quad {\left |{ {{\mathbf{v}}_{m,k}^{(i)~H}{{\mathbf{h}}_{m,mk}^{(i)}}} }\right |^{2}}{\alpha _{m,j}^{(i)}} \ge \Theta D, \tag{24a}
       \\&
        \qquad \quad\qquad
        \forall m,  k\in {\mathcal{L}_{m}} ,
       j \in \left \{{ k,k+1,\ldots,{L_{m}} }\right \} ,
       \\&\qquad\quad {\left |{ {{\mathbf{v}}_{m,k}^{(i)~H}{{\mathbf{h}}_{m,mk}^{(i)}}} }\right |^{2}}{\alpha _{m,j}^{(i)}} \ge D\left(2^{{R_{m,j}^{min}}}-1\right), \tag{24b}
       \\&
       \qquad \quad\qquad
        \forall m,  k\in {\mathcal{L}_{m}} ,
        j \in \left \{{ k,k+1,\ldots,{L_{m}} }\right \} ,
       \\&\qquad\quad {\sum \limits _{k = 1}^{L_{m}} {{\alpha _{m,k}^{(i)}}} \leq \frac {L_{m}}{K},} ~ \forall m, \tag{24c}
       \\&\qquad\quad {\alpha _{m,k}^{(i)}} \ge 0,~ \forall m,k \in {\mathcal{L}_{m}}, \tag{24d}
      \end{align*}
      where $D={{\sum \nolimits _{l = 1}^{j - 1} {{{\left |{ {{\mathbf{v}}_{m,k}^{(i)~H}{{\mathbf{h}}_{m,mk}^{(i)}}} }\right |}^{2}}{\alpha _{m,l}^{(i)}} + {{\|{ {{{\mathbf{v}}_{m,k}^{(i)}}} }\|}^{2}}\frac {1}{\rho }}}}$.

      We utilize a bisection searching method to find the optimal $\Theta$.
      Under given $\Theta$, $\mathcal {P}_{3}$ becomes a linear programming (LP) problem, which can be solved by standard LP tools to obtain the corresponding power allocation coefficient vector ${\boldsymbol {\alpha}} _{m}^{(i)}$.

  \item Step 2: To find the optimal user clustering mechanism, we would traverse all the user clustering combinations. Here, we adopt the exhaustive searching method.
\end{itemize}

The detailed process is shown in Algorithm \ref{erfensousuo}.
$SINR_{\min}^{\varphi}$ in Algorithm \ref{erfensousuo} denotes the smallest SINR of all the $M$ clusters under a given eMBB user clustering combination $\varphi \in \boldsymbol{\Lambda}$.
The computational complexity of step 1 and step 2 are ${\rm \mathcal {O}}\left ({2^K}\right )$ and ${\rm \mathcal {O}}\left ({ \frac{K!}{\prod_{m=1}^{M}{ L_{m} }!} }\right )$, respectively. Therefore, Algorithm \ref{erfensousuo} has the complexity of ${\rm \mathcal {O}}\left ({  2^K \frac{(K)!}{\prod_{m=1}^{M}{ L_{m} }!} }\right )$.

\begin{algorithm}[tbp]
\caption{Algorithm for user clustering.}
{\normalsize
\begin{algorithmic}[1] \label{erfensousuo}

\STATE \textbf{Initialize} the tolerance $\delta$ and the ${\small{SINR}}_{\min}=0$.

\STATE \textbf{For} each user combination $\varphi \in \boldsymbol{\Lambda}$ \textbf{do}

\STATE \hspace{3ex} Initialize the $SINR_{\min}^{\varphi}=0$ when the user group is
$\Lambda$.

\STATE \hspace{3ex} \textbf{For} each cluster $m \in {\mathcal{M}}$ \textbf{do}

\STATE \hspace{6ex} Initialize the lower bound of $\boldsymbol{SINR}_m$ as $\Theta_{LB}$,
the upper bound of $\boldsymbol{SINR}_m$ as $\Theta_{UB}$.

\STATE \hspace{6ex} \textbf{While} ${|\Theta_{UB}}-{\Theta_{LB}|}\geq \delta$ \textbf{do}

\STATE \hspace{9ex} Set $\Theta=\left({\Theta_{UB}}+{\Theta_{LB}}\right)/2$.

\STATE \hspace{9ex} Obtain ${\boldsymbol {\alpha}}_m$  by solving the linear problem $\mathcal {P}_{3}$.

\STATE \hspace{9ex} \textbf{If} $\mathcal {P}_{3}$ is feasible \textbf{then}

\STATE \hspace{12ex} Set $SINR_{\min}^m=\Theta$ and $\Theta_{LB}=\Theta$.

\STATE \hspace{9ex} \textbf{else}

\STATE \hspace{12ex} Set $\Theta_{UB}=\Theta$.

\STATE \hspace{9ex} \textbf{end If}

\STATE \hspace{6ex} \textbf{end While}

\STATE \hspace{6ex}Update $SINR_{\min}^{\varphi}\!=\!{\max}\!\left\{\!SINR_{\min}^{\varphi}, SINR_{\min}^m \! \right\}$.

\STATE \hspace{3ex} \textbf{end For}

\STATE \hspace{3ex} \textbf{If} $SINR_{\min}^{\varphi}>SINR_{\min}$ \textbf{then}

\STATE \hspace{6ex} Set $SINR_{\min}=SINR_{\min}^{\varphi}$ and ${\varphi}^{*}=\varphi$.

\STATE \hspace{3ex} \textbf{end if}

\STATE \textbf{end For}

\STATE \textbf{Output:}\hspace{1ex} $SINR_{\min}$ and the optimal user clustering combination ${\varphi}^{*}$.

\end{algorithmic}}
\end{algorithm}

\setcounter{equation}{24}
\section{Joint URLLC and eMBB Scheduling}\label{C}
In this section, we aim to find the optimal joint URLLC and eMBB scheduling scheme.

Intuitively, $\mathcal {P}_{1}$ is a non-convex optimization problem.
According to \eqref{sinr_embb} and \eqref{20}, the objective function relies on URLLC-eMBB punctured user selection and power assignment. Therefore, it follows two key problems: 1) which eMBB users to be punctured when URLLC users arrived? 2) how to assign transmission power to the users in each cluster?

As we can observed from \eqref{20}, the objective function is a coupled problem.
Considering the computational complexity, we decouple $\mathcal {P}_{1}$ into two subproblems, namely, the punctured user selection problem and the multi-cluster power allocation problem.
Thereafter, we propose an iterative algorithm, where the two subproblems are performed in an iterative way to obtain a solution for $\mathcal {P}_{1}$.
In this work, the punctured user selection problem can be formulated as a matching game. Under a punctured user selection scheme, the power allocation problem can be transformed as a convex problem by means of SCA and D.C. programming.

\subsection{The Power Allocation Problem}
Under a given punctured user selection scheme, the effective SINR for URLLC and eMBB users in the $m$-cluster at mini slot $i$ can be simplified as
\begin{equation}
\begin{split} \label{zongtisinr}
SINR_{m,k}^{e/u,(i)}={ {{{\left |{ {{\mathbf{m}}_{m,k}^{(i)~H}{{\mathbf{h}}_{m,mk}^{(i)}}} }\right |}^{2}}{\gamma _{m,k}^{(i)}}} \Big/ {\Big({{\|{ {{{\mathbf{m}}_{m,k}^{(i)}}} }\|}^{2}}\frac {1}{\rho }}}+\\
\!\sum \limits_{u \in \mathfrak{U_{1}} }
{{{\left |{ {{\mathbf{m}}_{m,k}^{(i)~H}{{\mathbf{h}}_{m,mk}^{(i)}}} }\right |}^{2}}{\beta _{m,u}^{(i)}}}
+ \sum \limits_{e \in \mathfrak{E_{1}} }{{\left |{ {{\mathbf{m}}_{m,k}^{(i)~H}{{\mathbf{h}}_{m,mk}^{(i)}}} }\right |}^{2}}{\alpha _{m,e}^{(i)}}
\!\Big),\\
\forall m,k \in \{{\mathcal {S}^{(i)}_{~\!e,m}}\cup {\mathcal {S}^{(i)}_{~\!u,m}}\},i,
\end{split}
\end{equation}
where ${\gamma _{m,k}^{(i)}}$ is the unified expression of ${\alpha _{m,k}^{(i)}}$ and ${\beta _{m,k}^{(i)}}$, ${\mathbf{m}}_{m,k}^{(i)}$ is the
unified expression of ${\mathbf{v}}_{m,k}^{(i)}$ and ${\mathbf{w}}_{m,k}^{(i)}$, the user sets $\mathfrak{S_{1}}$ and $\mathfrak{U_{1}}$ are respectively defined as
\begin{equation}
\mathfrak{S_{1}}={ \left \{{ {l \in {\mathcal{S}^{(i)}_{~\!e,m}}  \Big|
|{{\mathbf{m}}_{m,l}^{(i)~H}{{\mathbf{h}}_{m,ml}^{(i)}}}|
>
| {{\mathbf{m}}_{m,k}^{(i)~H}{{\mathbf{h}}_{m,mk}^{(i)}}}|
 } }\right \}},
\end{equation}
and
\begin{equation}
\mathfrak{U_{1}}={ \left \{{ {u \in {\mathcal{S}^{(i)}_{~\!u,m}}  \Big|
|{{\mathbf{m}}_{m,u}^{(i)~H}{{\mathbf{h}}_{m,mu}^{(i)}}}|
 >
 |{{\mathbf{m}}_{m,k}^{(i)~H}{{\mathbf{h}}_{m,mk}^{(i)}}}|
 } }\right \}}.
\end{equation}
%\begin{equation}
%\begin{aligned}
%\mathfrak{U_{1}}={ \left \{{ {u \in {\mathcal{S}^u_{m}(i)}  \mid
%{{\left |{ {{\mathbf{m}}_{m,u,i}^{H}{{\mathbf{h}}_{m,mu}^{(i)}}} }\right |}^{2}}
% >
% {{\left |{ {{\mathbf{m}}_{m,k,i}^{H}{{\mathbf{h}}_{m,mk}^{(i)}}} }\right |}^{2}}
% } }\right \}}.
%\end{aligned}
%\end{equation}

After puncturing, URLLC and eMBB users in cluster $m$ need to be rearranged based on the effective channel gains in a decreasing order for the implementation of SIC decoding.
For simplicity, we use the sorted order for subscripts in each formula below.
Hence, \eqref{zongtisinr} can be rewritten as
\begin{equation}
\begin{split} \label{lastSINR}
SINR_{m,k}^{e/u,(i)}
= \frac {{{{\left |{ {{\mathbf{m}}_{m,k}^{(i)~H}{{\mathbf{h}}_{m,mk}^{(i)}}} }\right |}^{2}}{\gamma _{m,k}^{(i)}}}}
{ { \!\sum \limits_{l=1 }^{k-1}
{{{\left |{ {{\mathbf{m}}_{m,k}^{(i)~H}{{\mathbf{h}}_{m,mk}^{(i)}}} }\right |}^{2}}{\gamma _{m,l}^{(i)}}
\!+\! {{\|{ {{{\mathbf{m}}_{m,k}^{(i)}}} }\|}^{2}}\frac {1}{\rho }}}},\\
\forall m,k \in \{{\mathcal {S}^{(i)}_{~\!e,m}}\cup {\mathcal {S}^{(i)}_{~\!u,m}}\},i.
\end{split}
\end{equation}
%where ${\gamma _{m,k}^{(i)}}$ is the unified expression of ${\alpha _{m,k}^{(i)}}$ and ${\beta _{m,k}^{(i)}}$.
The subscripts of ${\mathbf{m}}_{m,k}^{(i)}$, ${\mathbf{h}}_{m,mk}^{(i)}$, and ${\gamma _{m,k}^{(i)}}$ are all in order of decreasing ${\left |{ {{\mathbf{m}}_{m,k,i}^{H}{{\mathbf{h}}_{m,mk}^{(i)}}} }\right |}$.
%In addition, $k$ here and below refer to both the URLLC user and eMBB user.
Thus, the data rate for URLLC users ${r_{m,k}^{(i)}},\forall k \in \mathcal {S}^{(i)}_{~\!u,m}$ and eMBB users ${R_{m,k}^{(i)}}, \forall k \in \mathcal {S}^{(i)}_{~\!e,m}$ can be further converted based on \eqref{lastSINR}.

Taking into account the similarity of the problems among different mini slots, $\mathcal {P}_{1}$ can be rewritten as
\begin{align*}
&\mathcal {P}_{4}: \quad \max \limits _{\boldsymbol {\gamma}^{(i)}}  \sum \limits _{m = 1}^{M} \sum \limits _{k \in \mathcal{S}^{(i)}_{~\!\!e,m} }
 \log_2\left({1+SINR_{m,k}^{e/u,(i)}}\right), \tag{29}\label{maxmubiao}\\&\quad \, \mathrm{subject}~\mathrm{to}~\eqref{20a}, \eqref{20b},
%\\& \qquad\quad \frac {F_{m,k}^{(i)}}{r_{m,k}^{(i)}} \leq D_{m,k}^{(i), \max}, ~ \forall m, k \in \mathcal {S}^{(i)}_{~\!u,m}, i ,\tag{30a}\label{27a}
%\\& \qquad\quad {R_{m,k}^{(i)}}\ge {R_{m,k}^{\min }},~ \forall m,  k \in \mathcal {S}^{(i)}_{~\!e,m},i, \tag{30b} \label{27b}
\\& \qquad\quad {\gamma _{m,k}^{(i)}}>0, ~\forall m,k \in \{{\mathcal {S}_{~\!e,m}^{(i)}}\cup {\mathcal {S}_{~\!u,m}^{(i)}}\},i, \tag{29a}\label{27c}
\\& \qquad\quad \sum \nolimits _{k \in \{{\mathcal {S}_{~\!\!e,m}^{(i)}}\cup {\mathcal {S}_{~\!\!u,m}^{(i)}}\}}\!\!{\gamma _{m,k}^{(i)}}
 \leq \frac{{L_{m}}}{K}, ~ \forall m ,i, \tag{29b}\label{27d}
\end{align*}
%where $\boldsymbol {\gamma _{m}^{(i)}}=\left \{{ {\gamma _{m,1}^{(i)}}, \cdots , {\gamma _{m,L_m}^{(i)}} }\right \}$
where $\boldsymbol {\gamma}^{(i)}\in \mathbb {R}^{K \times 1}$
is the power allocation coefficient collections for the URLLC and eMBB users.

Nevertheless, it is still hard to tackle $\mathcal {P}_{4}$ due to the non-convexity of the objective function as well as constraints \eqref{20a} and \eqref{20b}. In what follows, we aim to solve $\mathcal{P}_{4}$ based on D.C. programming and SCA method.

\setcounter{equation}{29}
\subsubsection{ D.C. programming for the objective function}
The objective function \eqref{maxmubiao} in $\mathcal {P}_{4}$ can be equivalently transformed into a canonical form with the help of D.C. programming as \cite{DP}
\begin{equation}\label{28}
\underset {\boldsymbol{\gamma}^{(i)}}{ \mathop {\mathrm {minimize}}\nolimits } \;\;{G_{1}\left ({{\bm{\gamma}^{(i)}} }\right) - {G_{2}}\left ({{\bm{\gamma}^{(i)}} }\right)},
\end{equation}
where $G_{1}\left ({\boldsymbol{\gamma}}^{(i)}\right)$ and $G_{2}\left ({\boldsymbol{\gamma}}^{(i)}\right)$ are respectively given by
\begin{equation}
{G_{1}}\left ({\boldsymbol{\gamma}}^{(i)}\right)=-\sum \limits _{m = 1}^{M} {\sum \limits _{k \in {\mathcal{S}^{(i)}_{~\!\!e,m}}} {\log _{2}\left ({D_{1}^{m,k}\left ({\boldsymbol{\gamma}}^{(i)}\right) }\right)} },
\end{equation}
and
\begin{equation}
{G_{2}}\left ({\boldsymbol{\gamma}}^{(i)}\right)=-\sum \limits _{m = 1}^{M} {\sum \limits _{k \in {\mathcal{S}^{(i)}_{~\!\!e,m}}} {\log _{2}\left ({D_{2}^{m,k}\left ({\boldsymbol{\gamma}}^{(i)}\right) }\right)} },
\end{equation}
where $D_{1}^{m,k}\left ({\boldsymbol{\gamma}}^{(i)}\right)$ and $D_{2}^{m,k}\left ({\boldsymbol{\gamma}}^{(i)}\right)$ are respectively given by
\begin{equation}
D_{1}^{m,k}\left ({\boldsymbol{\gamma}}^{(i)}\right)=   \!\sum \limits_{l=1}^{k}
{{\left |{ {{\mathbf{m}}_{m,k}^{(i)~H}{{\mathbf{h}}_{m,mk}^{(i)}}} }\right |}^{2}}{\gamma _{m,l}^{(i)}}
\!+\! {{\|{ {{{\mathbf{m}}_{m,k}^{(i)}}} }\|}^{2}}\frac {1}{\rho },
\end{equation}
and
\begin{equation}
D_{2}^{m,k}\left ({\boldsymbol{\gamma}}^{(i)}\right)= { { \!\sum \limits_{l=1}^{k-1}
{{{\left |{ {{\mathbf{m}}_{m,k}^{(i)~H}{{\mathbf{h}}_{m,mk}^{(i)}}} }\right |}^{2}}{\gamma _{m,l}^{(i)}}
\!+\! {{\|{ {{{\mathbf{m}}_{m,k}^{(i)}}} }\|}^{2}}\frac {1}{\rho }}}}.
\end{equation}

Note that $G_{1}\left ({\boldsymbol{\gamma}}^{(i)}\right)$ and $G_{2}\left ({\boldsymbol{\gamma}}^{(i)}\right)$ are different convex functions with respect to ${\boldsymbol{\gamma}}^{(i)}$. For any feasible solution ${\boldsymbol{\gamma}}^{(i),p}$ in the $p$-th iteration, we can obtain a lower bound for $G_{2}\left ({\boldsymbol{\gamma}}^{(i)}\right)$, which is given by
\begin{equation}\label{g2}
 G_{2}\left ({\boldsymbol{\gamma}}^{(i)}\right)\!\ge\! G_{2}\left ({\boldsymbol{\gamma}}^{(i),p}\right)
 \!+\! {\nabla _{{\boldsymbol{\gamma}}^{(i)}}}{G_{2}}{\left ({\boldsymbol{\gamma}}^{(i),p}\right)^{\mathrm {T}}}\left ({{\boldsymbol{\gamma}}^{(i)} \!-\! {\boldsymbol{\gamma}}^{(i),p} }\right),
\end{equation}
where ${\nabla _{{\boldsymbol{\gamma}}^{(i)}}}{G_{2}}\left ({\boldsymbol{\gamma}}^{(i),p}\right) = \left \{{ {\frac {{\partial {G_{2}}\left ({\boldsymbol{\gamma}}^{(i)}\right)}}
{{\partial {\gamma_{m,k}^{(i)}}}}\left |{ {_{{{\boldsymbol{\gamma}}^{(i),p}}}} }\right.} }\right \} \in \mathbb {R}^{{ K} \times 1}$ denotes the gradient of ${G_{2}}\left ({\cdot }\right)$ with respect to ${\boldsymbol{\gamma}}^{(i)}$ and
\begin{equation}
\frac{\partial {G_{2}\left ({\boldsymbol{\gamma}}^{(i)}\right)}}{{\partial {\gamma_{m,k}^{(i)}}}} |{ {_{{{\boldsymbol{\gamma}}^{(i),p}}}} } = -\frac {1}{\ln (2) }\sum \limits_{k' \in \mathfrak{R_{2}}} {\frac
{{{\left |{ {{\mathbf{m}}_{m,k'}^{(i)~H}{{\mathbf{h}}_{m,mk'}^{(i)}}} }\right |}^{2}}}
{D_{2}^{m,k'}\left ({\boldsymbol{\gamma}}^{(i),p}\right)}},
\end{equation}
where
%\begin{equation}
%\begin{aligned}
%\mathfrak{R_{2}}\!=\!\left \{\!{ {{\mathcal{S}^u_{m}(i)} \cup {\mathcal{S}^e_{m}(i)} \mid
%{{\left |{ {{\mathbf{m}}_{m,k',i}^{H}{{\mathbf{h}}_{m,mk'}^{(i)}}} }\right |}^{2}}
% \!<\!
 %{{\left |{ {{\mathbf{m}}_{m,k,i}^{H}{{\mathbf{h}}_{m,mk}^{(i)}}} }\!\right |}^{2}}
% } }\right \}
%\end{aligned}
%\end{equation}
\begin{equation}
\mathfrak{R_{2}}\!=\!\left \{\!{ {k' \! \in \! {\mathcal{S}^{(i)}_{~\!u,m}}\! \cup \! {\mathcal{S}^{(i)}_{~\!e,m}} \Big|
|{{\mathbf{m}}_{m,k'}^{(i)~H}{{\mathbf{h}}_{m,mk'}^{(i)}}}|
\!<\!
|{{\mathbf{m}}_{m,k}^{(i)~H}{{\mathbf{h}}_{m,mk}^{(i)}}}|
 } }\right \}.
\end{equation}

Based on \eqref{g2}, we can obtain an upper bound for the objective function \eqref{28} as
\begin{multline}\label{36}
{G_{1}\!\left ({\boldsymbol{\gamma}}^{(i)}\right) \!-\! {G_{2}}\!\left ({\boldsymbol{\gamma}}^{(i)}\right) } \leq
 G_{1}\!\left ({\boldsymbol{\gamma}}^{(i)}\right) \!-\! {G_{2}}\!\left ({\boldsymbol{\gamma}}^{(i),p}\right)\! \\
 -\!{\nabla _{{\boldsymbol{\gamma}}^{(i)}}}{G_{2}}\!{\left ({\boldsymbol{\gamma}}^{(i),p}\right)^{\mathrm {T}}}\left ({{\boldsymbol{\gamma}}^{(i)} \!- \! {\boldsymbol{\gamma}}^{(i),p} }\right).
\end{multline}
Therefore, the objective function \eqref{28} is transferred to minimize the upper bound of \eqref{36}, which is a convex problem and can be solved by standard convex tools.

\subsubsection{ SCA for constraints \eqref{20a} and \eqref{20b}}
Substituting \eqref{urllcrate} and \eqref{embbrate}, \eqref{20a} and \eqref{20b} can be respectively rewritten as
\begin{multline}\label{urllc_yueshu}
\frac {F_{m,k}^{(i)}}{BD_{m,k}^{(i), \max} } \leq \log_2 (1+SINR_{m,k}^{e/u,(i)}) - \\
\sqrt {C_{m,k}^{(i)}} \frac {Q^{-1}(\epsilon)\log e}{\sqrt {b_{m,k}^{(i)}}},~
\forall m, k \in \mathcal {S}_{~\!u,m}^{(i)}, i,
\end{multline}
and
\begin{equation}\label{embb_yueshu}
 {R_{m,k}^{\min }}\le\log_2\left({1+SINR_{m,k}^{e/u,(i)}}\right),
 ~\forall m,k \in \mathcal {S}_{~\!e,m}^{(i)},i,
\end{equation}
where $\sqrt{C_{m,k}^{(i)}}$ and $\log_2\left({1+SINR_{m,k}^{e/u,(i)}}\right)$ can be respectively expressed as
\begin{multline}\label{last_C}
\sqrt{C_{m,k}^{(i)}}={
\Big(
{{{\left |{ {{\mathbf{m}}_{m,k}^{(i)~H}{{\mathbf{h}}_{m,mk}^{(i)}}} }\right |}^{2}}{\gamma _{m,k}^{(i)}}}
\Big)\!^{1\!/\!2}
\Big( {
\!2{{{\|{ {{{\mathbf{m}}_{m,k}^{(i)}}} }\|}^{2}}\frac {1}{\rho }}
 } }
\\ {
+\!{{\left |{ {{\mathbf{m}}_{m,k}^{(i)~H}{{\mathbf{h}}_{m,mk}^{(i)}}} }\right |}^{2}}{\gamma _{m,k}^{(i)}}+\!2\!\sum \limits_{l=1 }^{k-1}
{{\left |{ {{\mathbf{m}}_{m,k}^{(i)~H}{{\mathbf{h}}_{m,mk}^{(i)}}} }\right |}^{2}}{\gamma _{m,l}^{(i)}}
}
\Big)\!^{1\!/\!2}\! \Big/
\\
\Big(
{2\!\sum \limits_{l=1 }^{k-1}
{{{\left |{ {{\mathbf{m}}_{m,k}^{(i)~H}{{\mathbf{h}}_{m,mk}^{(i)}}} }\right |}^{2}}{\gamma _{m,l}^{(i)}}
\!+\!{{{\left |{ {{\mathbf{m}}_{m,k}^{(i)~H}{{\mathbf{h}}_{m,mk}^{(i)}}} }\right |}^{2}}{\gamma _{m,k}^{(i)}}}
}}
\Big),
\end{multline}
and
\begin{equation}
\begin{split}
&\log_2\left({1+SINR_{m,k}^{e/u,(i)}}\right)\\
&= \log_2\left({{ \!\sum \limits_{l=1 }^{k}
{{{\left |{ {{\mathbf{m}}_{m,k}^{(i)~H}{{\mathbf{h}}_{m,mk}^{(i)}}} }\right |}^{2}}{\gamma _{m,l}^{(i)}}
\!
+\! {{\|{ {{{\mathbf{m}}_{m,k}^{(i)}}} }\|}^{2}}\frac {1}{\rho }}}}\right) \\
&~~~-
\log_2\left({{ \!\sum \limits_{l=1 }^{k-1}
{{{\left |{ {{\mathbf{m}}_{m,k}^{(i)~H}{{\mathbf{h}}_{m,mk}^{(i)}}} }\right |}^{2}}{\gamma _{m,l}^{(i)}}
\!+\! {{\|{ {{{\mathbf{m}}_{m,k}^{(i)}}} }\|}^{2}}\frac {1}{\rho }}}}\right).
\end{split}
\end{equation}

Intuitively, \eqref{embb_yueshu} is concave. By analyzing \eqref{channel_dis}, we can verify that $\sqrt{C_{m,k}^{(i)}}$ is concave with respect to $1+SINR_{m,k}^{e/u,(i)}>1$.
On this basis, we can employ the SCA mechanism to deal with the non-convex constraints \eqref{20a} and \eqref{20b} \cite{jianhua-tang-urllc-embb-slice}. Specifically, SCA performs a locally tight approximation of the original problem at each iteration to produce a tight convex constrain sets. According to \cite{jianhua-tang-urllc-embb-slice}, the approximation functions for both \eqref{20a} and \eqref{20b} should hold the following conditions: 1) they are continuous, consistent, and convex; 2) they are upper-bounds of the original functions.

In what follows, we would apply SCA to \eqref{urllc_yueshu} and \eqref{embb_yueshu} to obtain their upper bounds.
Let's define
\begin{multline}
\hspace{-1em}H_1^{m,k}({\bm{\gamma}^{(i)}})=\log_2\left({{ \!\sum \limits_{l=1 }^{k-1}
{{{\left |{ {{\mathbf{m}}_{m,k}^{(i)~H}{{\mathbf{h}}_{m,mk}^{(i)}}} }\right |}^{2}}{\gamma _{m,l}^{(i)}}
\!+\! {{\|{ {{{\mathbf{m}}_{m,k}^{(i)}}} }\|}^{2}}\frac {1}{\rho }}}}\right), \\
\forall m,k \in \mathcal {S}_{~\!u,m}^{(i)},i.
\end{multline}
Applying SCA to $H_1^{m,k}(\boldsymbol{\gamma}^{(i)})$, we have
\begin{multline}\label{H1}
H_1^{m,k}\left (\boldsymbol{\gamma}^{(i)}\right)\!\le\! H_1^{m,k}\left (\boldsymbol{\gamma}^{(i),p}\right)
 \!+\\
 \! {\nabla_{\bm{\gamma}^{(i)}}}{H}_1^{m,k}\!{\left (\boldsymbol{\gamma}^{(i),p}\right)\!^{\mathrm {T}}}\left(\boldsymbol{\gamma}^{(i)} -\boldsymbol{\gamma}^{(i),p}\right),
\end{multline}
where we denote the gradient of $H_1^{m,k}\left(\cdot\right)$ with respect to ${\bm{\gamma}^{(i)}}$ as ${\nabla _{\bm{\gamma}^{(i)}}}{H}_1^{m,k}\left ({{\bm{\gamma}}^{(i),p} }\right) \!=\! \left \{{ {\frac {{\partial {H}_1^{m,k}\left ({{\bm{\gamma}^{(i)}} }\right)}}
{{\partial {\gamma_{m',k'}^{(i)}}}}\left |{ {_{{{\bm{\gamma}^{(i),p}}}}} }\right.} }\right \}_{k' \in \{{\mathcal {S}_{~\!\!e,m'}^{(i)}}\cup {\mathcal {S}_{~\!\!u,m'}^{(i)}}\}}^{m' \in \mathcal{M}}$ and $\frac{\partial {H_1^{m,k}\left ({{\bm{\gamma}^{(i)}} }\right)}}{{\partial {\gamma_{m',k'}^{(i)}}}} |{ {_{{{\bm{\gamma}^{(i),p}}}}} }$ is defined in \eqref{43}.

\begin{figure*}[ht]
\normalsize
\setcounter{equation}{44}
\setcounter{mytempeqncnt1}{\value{equation}}
 \begin{equation}
\begin{aligned} \label{43}
\frac{\partial {H_1^{m,k}\left ({{\bm{\gamma}^{(i)}}}\right)}}{{\partial {\gamma_{m',k'}^{(i)}}}} |{ {_{{{\bm{\gamma}^{(i),p}}}}} } =
\left\{\begin{array}{rcl} \frac {{{{\left |{ {{\mathbf{m}}_{m,k}^{(i)~H}{{\mathbf{h}}_{m,mk}^{(i)}}} }\right |}^{2}}}}{\ln (2) \left({{ \!\sum \limits_{l=1 }^{k-1}
{{{\left |{ {{\mathbf{m}}_{m,k}^{(i)~H}{{\mathbf{h}}_{m,mk}^{(i)}}} }\right |}^{2}}{\gamma_{m',l}^{(i),p}}
\!+\! {{\|{ {{{\mathbf{m}}_{m,k}^{(i)}}} }\|}^{2}}\frac {1}{\rho }}}}\right)}, \quad &m'= m,k'=1,2,\cdots k-1,
\\0, & \mathrm{otherwise}. \end{array}\right.
\end{aligned}
\end{equation}
\setcounter{equation}{\value{mytempeqncnt1}}
\end{figure*}
\begin{figure*}[ht]
\vspace{-0.2cm}
%\hrulefill
\normalsize
\setcounter{equation}{46}
\setcounter{mytempeqncnt1}{\value{equation}}
\begin{equation}
\begin{aligned}\label{h2}
\frac{\partial {H_2^{m,k}\left ({{\bm{\gamma}^{(i)}}}\right)}}{{\partial {\gamma_{m',k'}^{(i)}}}} |{ {_{{{\bm{\gamma}^{(i),p}}}}}} =
\left\{\begin{array}{rcl} \frac {S_{1}{{{\left |{ {{\mathbf{m}}_{m,k}^{(i)~H}{{\mathbf{h}}_{m,mk}^{(i)}}} }\right |}^{2}}(S_3-2S_2^2)}}{S_2S_3^2},
\quad &m'= m,k'=1,2,\cdots k-1,\\
\frac{{{\left |{ {{\mathbf{m}}_{m,k}^{(i)~H}{{\mathbf{h}}_{m,mk}^{(i)}}} }\right |}^{2}}S_2}{2S_1S_3}-\frac{S_{1}{{{\left |{ {{\mathbf{m}}_{m,k}^{(i)~H}{{\mathbf{h}}_{m,mk}^{(i)}}} }\right |}^{2}}(S_3-2S_2^2)}}{2S_2S_3},
\quad &m'= m,k'=k,
\\0, & \mathrm{otherwise}. \end{array}\right.
\end{aligned}
\end{equation}
\setcounter{equation}{\value{mytempeqncnt1}}
\end{figure*}

\setcounter{equation}{45}
Similarly, applying SCA to ${H_2^{m,k}}=\sqrt{C_{m,k}^{(i)}},\forall m,k \in \mathcal {S}_{~\!u,m}^{(i)},i$, we have
\begin{multline}\label{H2}
 H_2^{m,k}\!\left ({{\bm{\gamma}^{(i)}} }\right)\!\le\! H_2^{m,k}\left ({{\bm{\gamma}^{(i),p}} }\right)
 \!+\\
 \! {\nabla _{\bm{\gamma}^{(i)}}}{H}_2^{m,k}\!{\left ({{\bm{\gamma}^{(i),p}} }\right)\!^{\mathrm {T}}}\!\left ({{\bm{\gamma}^{(i)}} \!-\! {\bm{\gamma}^{(i),p}} }\right),
\end{multline}
where we denote the gradient of ${H_2^{m,k}}\left ({\cdot }\right)$ with respect to ${\bm{\gamma}^{(i)}}$ as  ${\nabla _{\bm{\gamma}^{(i)}}}{H}_2^{m,k}\left ({{\bm{\gamma}^{(i),p}} }\right) = \left \{{ {\frac {{\partial {H}_1^{m,k}\left ({{\bm{\gamma}^{(i)}} }\right)}}
{{\partial {\gamma_{m',k'}^{(i)}}}}\!\left |{ {_{{{\bm{\gamma}^{(i),p}}}}} }\right.} }\right \}_{k' \in \{{\mathcal {S}_{~\!\!e,m'}^{(i)}}\cup {\mathcal {S}_{~\!\!u,m'}^{(i)}}\}}^{m' \in \mathcal{M}}$ and $\frac{\partial {H_2^{m,k}\left ({{\bm{\gamma}^{(i)}} }\right)}}{{\partial {\gamma_{m',k'}^{(i)}}}} \!|\!{ {_{{{\bm{\gamma}^{(i),p}}}}} }$ is defined in \eqref{h2}.
\setcounter{equation}{47}
Here, ${S_{1}}$, ${S_{2}}$, and ${S_{3}}$ are respectively given by
\begin{equation}
{S_{1}}= \Big({{{\left |{ {{\mathbf{m}}_{m,k}^{(i)~H}{{\mathbf{h}}_{m,mk}^{(i)}}} }\right |}^{2}}{\gamma _{m,k}^{(i),p}}}\Big)\!^{1\!/\!2},
\end{equation}
\begin{multline}\label{s2}
{S_{2}}={
\Big(
{
\sum \limits_{l=1 }^{k-1}
{{\left |{ {{\mathbf{m}}_{m,k}^{(i)~H}{{\mathbf{h}}_{m,mk}^{(i)}}} }\right |}^{2}}{\gamma _{m,l}^{(i),p}}\!+\!
2{{{\|{ {{{\mathbf{m}}_{m,k}^{(i)}}} }\|}^{2}}\frac {1}{\rho }}
}
}\\
+{{{{\left |{ {{\mathbf{m}}_{m,k}^{(i)~H}{{\mathbf{h}}_{m,mk}^{(i)}}} }\right |}^{2}}{\gamma _{m,k}^{(i),p}}}
}
\Big)\!^{1\!/\!2},
\end{multline}
\begin{equation}
{S_{3}}=
{2\!\sum \limits_{l=1 }^{k-1}
{{{\left |{ {{\mathbf{m}}_{m,k}^{(i)~H}{{\mathbf{h}}_{m,mk}^{(i)}}} }\right |}^{2}}{\gamma _{m,l}^{(i),p}}
\!+\!{{{\left |\!{ {{\mathbf{m}}_{m,k}^{(i)~H}\!{{\mathbf{h}}_{m,mk}^{(i)}}} }\right |}^{2}}{\gamma _{m,k}^{(i),p}}}
}}.
\end{equation}

Therefore, constraints \eqref{20a} at the $p$-th iteration can be approximated as
\begin{multline}\label{urllc_zhuanhuan}
\frac {F_{m,k}^{(i)}}{BD_{m,k}^{(i), \max} }  \leq \log_2\left({{ \!\sum \limits_{l=1 }^{k}
{{{\left |{ {{\mathbf{m}}_{m,k}^{(i)~H}{{\mathbf{h}}_{m,mk}^{(i)}}} }\right |}^{2}}{\gamma _{m,l}^{(i)}}
\!
+\! {{\|{ {{{\mathbf{m}}_{m,k}^{(i)}}} }\|}^{2}}\frac {1}{\rho }}}}\right)\\
-{\frac {Q^{-1}(\epsilon)\log e}{\sqrt {b_{m,k}^{(i)}}}
\Big(
{
H_2^{m,k}\left ({\boldsymbol{\gamma}}^{(i),p}\right)
+ {\nabla _{{\boldsymbol{\gamma}}^{(i)}}}{H}_2^{m,k}{\left ({\boldsymbol{\gamma}}^{(i),p}\right)^{\mathrm {T}}}\cdot
}
}\\
\left ({{\boldsymbol{\gamma}}^{(i)} - {\boldsymbol{\gamma}}^{(i),p} }\right)
\Big)
 -{\nabla _{{\boldsymbol{\gamma}}^{(i)}}}{H}_1^{m,k}{\left ({\boldsymbol{\gamma}}^{(i),p}\right)^{\mathrm {T}}}\left ({{\boldsymbol{\gamma}}^{(i)} - {\boldsymbol{\gamma}}^{(i),p} }\right)
 \\
\quad\quad\quad\quad\quad\quad\quad\quad -H_1^{m,k}\left ({\boldsymbol{\gamma}}^{(i),p}\right), ~
 \forall m,k \in \mathcal {S}_{~\!u,m}^{(i)}, i,
\end{multline}
and constraints \eqref{20b} at the $p$-th iteration can be approximated as
\begin{multline}\label{embb_zhuanhuan}
 {R_{m,k}^{\min }}\le
 \log_2\left({{ \!\sum \limits_{l=1 }^{k}
{{{\left |{ {{\mathbf{m}}_{m,k}^{(i)~H}{{\mathbf{h}}_{m,mk}^{(i)}}} }\right |}^{2}}{\gamma _{m,l}^{(i)}}
\!
+\! {{\|{ {{{\mathbf{m}}_{m,k}^{(i)}}} }\|}^{2}}\frac {1}{\rho }}}}\right)\\
-H_1^{m,k}\left ({{\bm{\gamma}^{(i),p}} }\right)
  -{\nabla _{\bm{\gamma}^{i}}}{H}_1^{m,k}{\left ({{\bm{\gamma}^{(i),p}} }\right)^{\mathrm {T}}}\left ({{\bm{\gamma}^{(i)}} - {\bm{\gamma}^{(i),p}} }\right), \\
\forall m,k \in \mathcal {S}_{~\!e,m}^{(i)}, i.
\end{multline}

In this way, we can transform $\mathcal {P}_{4}$ as
\begin{multline}
\mathcal {P}_{5}: \quad \underset {\boldsymbol{\gamma}^{(i)}}{ \mathop {\mathrm {minimize}}\nolimits } \;\;G_{1}\left ({{\bm{\gamma}^{(i)}} }\right) - {G_{2}}\left ({{\bm{\gamma}^{(i),p}} }\right)\\
-{\nabla _{\bm{\gamma}^{(i)}}}{G_{2}}{\left ({{\bm{\gamma}^{(i),p}} }\right)^{\mathrm {T}}}\left ({{\bm{\gamma}^{(i)}} - {\bm{\gamma}^{(i),p}} }\right),
\end{multline}
subject to \eqref{27c}, \eqref{27d}, \eqref{urllc_zhuanhuan}, and \eqref{embb_zhuanhuan}.

Obviously, $\mathcal {P}_{5}$ is a convex problem.
An iteration algorithm is developed to solve $\mathcal {P}_{5}$, which is shown in Algorithm \ref{power}. Specifically, in the $p$-th iteration, the updated solution ${\bm{\gamma}}^{(i),p}$ is obtained by solving $\mathcal {P}_{5}$ with ${\bm{\gamma}}^{(i),p-1}$. The algorithm will terminate when the change of optimal objective function value becomes smaller than a given convergence tolerance, i.e., $|W^{p}\left ({{\bm{\gamma}^{(i)}} }\right)- W^{p-1}\left ({{\bm{\gamma}^{(i)}} }\right)| \le \epsilon$. Here, $\epsilon>0$ is a small constant, and
\begin{multline}\label{W}
W^{p}\left ({{\bm{\gamma}^{(i)}} }\right)=G_{1}\!\left ({{\bm{\gamma}^{(i)}} }\right) \!-\! {G_{2}}\!\left ({{\bm{\gamma}^{(i),p}} }\right)\!-\\
\!{\nabla _{\bm{\gamma}^{(i)}}}{G_{2}}\!{\left ({{\bm{\gamma}^{(i),p}} }\right)^{\mathrm {T}}}\left ({{\bm{\gamma}^{(i)}} \!- \! {\bm{\gamma}^{(i),p}} }\right).
\end{multline}

\begin{algorithm}[ht]
\caption{Joint URLLC and eMBB scheduling algorithm.}
\label{power}
{\normalsize
\begin{algorithmic}[1]

\STATE \textbf{Initialize} the convergence tolerance $\epsilon$.

\STATE Set the mini slot index $i=1$.

\STATE \textbf{Do}

\STATE \hspace{3ex} Define $\mathcal{S}^{(i)}_{~\!u}$ as the set of arrival URLLC users at mini slot $i$.

\STATE \hspace{3ex} The selection of the punctured eMBB users for arriving URLLC users is determined by Algorithm \ref{GS_suanfa}.

\STATE \hspace{3ex} Set the iteration index $p=1$.

\STATE \hspace{3ex} Initialize the feasible solution ${\boldsymbol{\gamma}}^{(i),1}$.

\STATE \hspace{3ex} \textbf{While} $|W^{p}\left ({{\bm{\gamma}^{(i)}} }\right)- W^{p-1}\left ({{\bm{\gamma}^{(i)}} }\right)| \geq \epsilon$ \textbf{do}

\STATE \hspace{6ex} Solving problem $\mathcal {P}_{5}$ under given ${\boldsymbol{\gamma}}^{(i),p-1}$ to obtain

the power allocation ${\boldsymbol{\gamma}}^{(i),p}$.

\STATE \hspace{6ex} Set $p=p+1$.

\STATE \hspace{3ex} \textbf{end While}

\STATE \hspace{3ex} Return the solution ${\boldsymbol{\gamma}}^{(i),*}={\boldsymbol{\gamma}}^{(i),p}$.

\STATE \hspace{3ex} Set $i=i+1$.

\STATE \textbf{Until} $i=Q+1$.
\end{algorithmic}}
\end{algorithm}

Since $\mathcal{P}_5$ is a convex formulation with linear matrix inequality (LMI) and second order cone (SOC) constraints, the computational complexity of Algorithm \ref{power} consists of two parts \cite{complexity}:
\begin{enumerate}
  \item \emph{Iteration Complexity}: The number of iterations required to reach $|W^{p}\left ({{\bm{\gamma}^{(i)}} }\right)- W^{p-1}\left ({{\bm{\gamma}^{(i)}} }\right)| < \epsilon$ is on the order of $\sqrt{\beta}$, where $\beta=\sum_{j=1}^{J}k^L_j+2m$, $J$ is the number of LMI constraints, $k^L_j$ means the dimension of the $j$-th LMI constraint, and $m$ denotes the number of SOC constraints. Here, we can find that $J=K+M$, $k^L_j=1, \forall j$, and $m=K$, i.e., $\sqrt{\beta}=\sqrt{3K+M}$.
  \item \emph{Per-Iteration Complexity Cost}: In each iteration, the computation cost is dominated by: (a) the formation of the coefficient matrix, which can be formulated as
      $C_{form}=n\sum_{j=1}^J (k_j^L)^3+n^2\sum_{j=1}^J (k_j^L)^2+n\sum_{j=1}^m (k_j^S)^2$,
      where $n$ is the total number of optimization parameters, $k^S_j$ is the dimension of the $j$-th SOC constraint;
      (b) the factorization of the coefficient matrix, which is given by $C_{fact}=n^3$.
      For Algorithm 2, we can find that $n=K$ and $k^S_j=N+1, \forall j$. Hence, the per-iteration complexity cost is $C_{form}+C_{fact}=K(K+M)+K^2(K+M)+K^2(N+1)^2+K^3$.
\end{enumerate}
By combining the above two parts, the computational complexity of Algorithm \ref{power} is on the order of
$\sqrt{\beta}\left(C_{form}+C_{fact}\right)$, which is derived as $\mathcal{O}\left(\sqrt{3K+M}\left(2K^3+K^2\left(N^2+2N+M+2\right)+KM\right)\right)$.
Moreover, the convergence of the proposed algorithm is proved in Appendix \ref{proofdc}.

\subsection{The Punctured User Selection Problem}
In this subsection, we investigate the punctured user selection problem. The user selection is represented by a binary variable $x _{m,k,n}^{(i)}$, where $x _{m,k,n}^{(i)}=1$ represents eMBB user $k$ in the $m$-th cluster is punctured by URLLC user $n$, and $x _{m,k,n}^{(i)}=0$ otherwise.
Matching theory \cite{match2}
provides a framework to tackle the above problem through pairing URLLC and eMBB players in two distinct sets based on each player's individual preference.
Similar with \cite{chenlaoshi}, we apply the Gale-Shapely (GS) algorithm with incomplete preference list here to match URLLC and eMBB pairs.

The GS algorithm can be used to find a stable one-to-one matching between men and women, which has been respectively regarded as URLLC and eMBB users in our work.
The basic idea of GS algorithm is that one gender have a sequence of ``proposals'' to the other sex (i.e. women).
When a woman receives a proposal, she rejects if she already holds a better proposal, and otherwise agrees to hold it for consideration.
In our matching model, the punctured user selection problem occurs in each mini slot.
At mini slot $i$ if BS allocates resources to URLLC $u_n \in \mathcal{S}^{(i)}_{~\!u}$ that is already hold by eMBB user $ e_{mk}\in {\mathcal{L}}$, then URLLC user $u_n$ is said to be matched with eMBB user $e_{mk}$ and form a matching pair $(u_n,e_{mk})$. Therefore, in this work, the matching is an assignment of URLLC users in $\mathcal{S}^{(i)}_{~\!u}$ to eMBB users ${\mathcal{L}}$, which can be defined as follows \cite{a-matching-embb-urllc-coexistence}.
\begin{definition}
\label{mmo}
A matching $\Omega$ between $\mathcal{S}^{(i)}_{~\!u}$ and ${\mathcal{L}}$ is a mapping from the set $\mathcal{S}^{(i)}_{~\!u}\cup {\mathcal{L}}$ to the set of all subsets of $\mathcal{S}^{(i)}_{~\!u} \cup {\mathcal{L}}$ such that for $\forall u_n \in \mathcal{S}^{(i)}_{~\!u}$ and $\forall e_{mk} \in {\mathcal{L}}$: (i) $\Omega(u_n) \in {\mathcal{L}}$ and $\Omega(e_{mk}) \in \mathcal{S}^{(i)}_{~\!u}$, (ii) $\left |{\Omega(u_n)}\right | =1$, (iii) $\left |{\Omega(e_{mk})} \right | \leq 1$.
\end{definition}

Definition \ref{mmo} illustrates that an URLLC user $u_n$ must be matched with one eMBB user from ${\mathcal{L}}$, while one eMBB user $e_{mk}$ can only be matched at most one URLLC user from $\mathcal{S}^{(i)}_{~\!u}$ at any mini slot.

The matching is performed based on the players' preference lists.
% In our model, the two sides, URLLC and eMBB users, would build their preference lists by employing available information at each side.
% To better describe the interaction between the two sides, we need to investigate how URLLC users puncture eMBB users and define preferences among URLLC and eMBB users. The details are as follow.
The preference list for eMBB users is defined based on a proposed concept of $only$ $one$ $URLLC$ $rate$ $(OOUR)$ of a certain URLLC user $u_n$.
$OOUR$ is the throughput if the BS only accept one URLLC user $u_n$, which punctures the eMBB user $k$ in the $m$-th cluster $e_{mk}$ and reject all other URLLC users at mini slot $i$, denoted as $R_n^{mk}$.
$R_n^{mk}$ can be obtained by solving problem $\mathcal {P}_{5}$ under a given matching matrix.
Thus, we define the preference relationship for eMBB user $e_{mk}$ between URLLC user $u_n$ and $u_{n'}$ in Definition \ref{embb_prefer}.
\begin{definition}\label{embb_prefer}
At mini slot $i$, eMBB user $e_{mk}$ prefers to choose URLLC user $u_n$ than URLLC user $u_{n'}$, if $R_n^{mk}>R_{n'}^{mk}$, denoted by $u_n\succ_{e_{mk}}{u_{n'}}$, for $ u_n\in \mathcal {S}^{(i)}_{~\!u}$, $ u_n'\in \mathcal {S}^{(i)}_{~\!u}$, $e_{mk} \in \mathcal{L}$.
%k \in \mathcal{E}_m, m\in \mathcal{L}$.
\end{definition}

\begin{algorithm}[H]
\caption{The GS algorithm for URLLC and eMBB users.}
\label{GS_suanfa}
{\normalsize
\begin{algorithmic}[1] \label{GS}

\STATE Set up eMBB users' preference lists as $\mathcal{PL}_{mk}^e, \forall e_{mk}\in {\mathcal{L}}$ based on Definition \ref{embb_prefer}.

\STATE Set up URLLC users' preference lists as $\mathcal{PL}_{n}^u, \forall u_n\in\mathcal {S}^{(i)}_{~\!u}$ based on Definition \ref{urllc-1} and Definition \ref{urllc-2}.

\STATE Define $\mathcal{M}_c=\mathcal {S}^{(i)}_{~\!u}$ as the set of unmatched URLLC users at mini slot $i$.

\STATE \textbf{While} $\mathcal{M}_c$ is nonempty \textbf{do}

\STATE \hspace{3ex} URLLC user $u_n$ selects the eMBB user $e_{mk}$ that
 locates first in its preference list $\mathcal{PL}_n^u$.

\STATE \hspace{3ex} \textbf{If} $e_{mk}$ has received a proposal  $u_{n'}$ in its preference list, \textbf{then}

\STATE \hspace{6ex} \textbf{If} $e_{mk}$ prefers $u_{n'}$, \textbf{then}

\STATE \hspace{9ex} $e_{mk}$ holds $u_{n'}$ and rejects $u_n$.

\STATE \hspace{9ex} Remove $u_{n'}$ from $\mathcal{M}_c$.

\STATE \hspace{9ex} Add $u_n$ into $\mathcal{M}_c$.

\STATE \hspace{6ex} \textbf{else}

\STATE \hspace{9ex} $e_{mk}$ rejects $u_{n'}$ and accept $u_{n}$.

\STATE \hspace{9ex} Remove $u_{n}$ from $\mathcal{M}_c$.

\STATE \hspace{9ex} Add $u_n'$ into $\mathcal{M}_c$.

\STATE \hspace{6ex} \textbf{end If}

\STATE \hspace{3ex} \textbf{else}

\STATE \hspace{6ex} $e_{mk}$ accept $u_{n}$.

\STATE \hspace{6ex} Remove $u_{n}$ from $\mathcal{M}_c$.

\STATE \hspace{3ex} \textbf{end If}

\STATE \textbf{End while}
\end{algorithmic}}
\end{algorithm}

When considering URLLC users' preference list, we similarly take $R_n^{mk}$ as an evaluation indicator.
To satisfy the QoS requirements of both URLLC and eMBB users, we introduce  $\overline{x}_{mk}^{(i)}, \forall m,k,i$ as another measure indicator to represent the number of mini slots that eMBB user $e_{mk}$ punctured by URLLC users from mini slot $1$ to mini slot $(i-1)$.
Thereafter, we define the preference relationship between URLLC user $u_n$ to eMBB user $e_{mk}$ and $e_{m'k'}$ in Definition \ref{urllc-1} and Definition \ref{urllc-2}.
\begin{definition}\label{urllc-1}
At mini slot $i$, URLLC user $u_n$ prefers to replace eMBB user $e_{mk}$ than eMBB user $e_{m'k'}$, if $\overline{x}_{mk}^{(i)}<\overline{x}_{m'k'}^{(i)}$, denoted by $e_{mk}\succ_{u_n}{e_{m'k'}}$, for $u_n\in \mathcal {S}^{(i)}_{~\!u}$, $e_{mk} \in \mathcal{L}$, $e_{m'k'} \in \mathcal{L}$.
\end{definition}
\begin{definition}\label{urllc-2}
At mini slot $i$, URLLC user $u_n$ prefers to replace eMBB user $e_{mk}$ than eMBB user $e_{m'k'}$, if $\overline{x}_{mk}^{(i)}=\overline{x}_{m'k'}^{(i)}$ and $R_n^{mk}>R_n^{m'k'}$, denoted by $e_{mk}\succ_{u_n}{e_{m'k'}}$, for $ u_n\in \mathcal {S}^{(i)}_{~\!u}$, $e_{mk} \in \mathcal{L}$, $e_{m'k'} \in \mathcal{L}$.
\end{definition}

\par
The preference list of eMBB user $e_{mk} \in \mathcal{L}$ over URLLC users $u_n \in \mathcal {S}^{(i)}_{~\!u}$, denoted by $\mathcal{PL}_{mk}^e$, is ranked by Definition \ref{embb_prefer} in a descending order.
Similarly URLLC user $u_n$'s preference list $\mathcal{PL}_n^c$ over $e_{mk}$
is ranked by Definition \ref{urllc-1} and Definition \ref{urllc-2} in a descending order.
Now we define stable matching in Definition \ref{stable}.
\begin{definition}\label{stable}
A matching $\Omega$ is stable, if there exists no blocking pair $(u_n,e_{mk})$, such that $e_{mk}\succ_{u_n}{\Omega(u_n)}$ and $u_n\succ_{e_{mk}}{\Omega(e_{mk})}$, where ${\Omega(u_n)}$ represents URLLC user $u_n$'s partner in $\Omega$ and ${\Omega(e_{mk})}$ represents eMBB user $e_{mk}$'s partner in $\Omega$.
\end{definition}
 Algorithm \ref{GS} presents how to use the GS algorithm to find a stable matching between URLLC and eMBB users,
 which has the complexity of ${\rm \mathcal{O}}\left (  { \left |{{  \mathcal{S}^{(i)}_{~\!u}}}\right |}^{2}  \right )$.

As shown in Algorithm \ref{GS_suanfa}, the exchange operations occur only if the utilities of URLLC and eMBB users are strictly improved.
After searching all the possible swaps, the matching stage
terminates, which means there has no swap operation to further improve the utilities of URLLC and eMBB users. Therefore, Algorithm \ref{GS} is also stable.

\section{Simulation Results}
\label{D}
\begin{figure}
  \centering
  % Requires \usepackage{graphicx}width=9cm, height=6.5cm
  \includegraphics[width=0.45\textwidth]{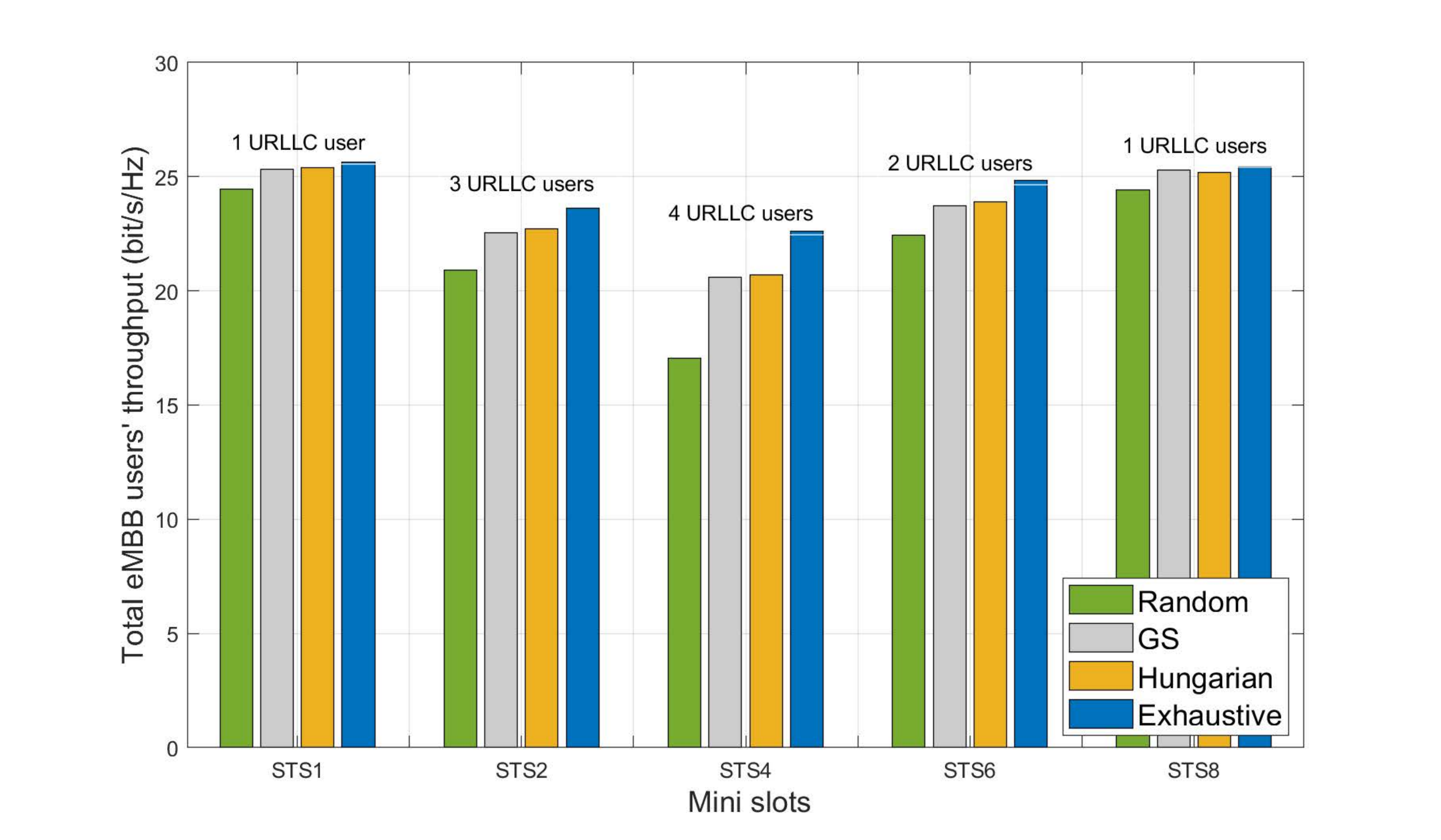}\\
  \caption{Performance comparison under different punctured schemes.}\label{fig:2}
\end{figure}
In this section, we would evaluate the performance of the proposed URLLC and eMBB coexistence network.
We consider $9$ eMBB users and $3$ clusters in the system. For simplicity, we let $4$, $3$, and $2$ eMBB users respectively exist in each cluster at the beginning of the transmission period. In each transmission period, there are $8$ mini slots. Furthermore, we let either the BS, the eMBB user, or the URLLC user equips $M=3$ antennas. In addition, in this work, we consider rayleigh fading channels.
Each URLLC user's packet length $F_{m,n}^{(i)}$ and blocklength $b_{m,n}^{(i)}$ are set as $500$ bytes and $168$, respectively \cite{jianhua-tang-urllc-embb-slice}. The reliability requirements of URLLC users are set as $99.999\%$. The tolerance $\delta$ and $\epsilon$ are set as the same, i.e., $10^{-6}$.
We assume the QoS requirements of eMBB users are the same, which is set as $R_{m,k}^{\min}=1$ bit/s/Hz. The delay requirements of URLLC users are the same, which is set as $D_{m,n}^{(i), \max}=0.7$ ms. In addition, the transmit SNR $\rho$ is set as $30$ dB.
For simplicity, we further assume there has at most one URLLC user arrives at a mini slot.

\par
In Fig. \ref{fig:2}, we present the punctured eMBB performance under different user selection algorithms, i.e.,  random searching method, GS algorithm, Hungarian algorithm, and exhaustive searching method.
Due to the computational complexity of the exhaustive searching method, we consider either the BS or the eMBB user has $3$ antennas.
Moreover, we assume there are $1$, $3$, $4$, $2$, and $1$ URLLC users arriving in mini slot $1$, $2$, $4$, $6$, and $8$, respectively.
From this figure, the performance gap between GS algorithm and exhaustive searching method would be less than $10\%$.
Moreover, compared with the random searching method, the GS algorithm has a significant performance improvement.
Since the number of punctured eMBB users increases with the number of arrival URLLC users, the performance gaps among the proposed $3$ schemes are the most apparent in STS4. The performance of the Hungarian algorithm is similar to the GS algorithm. However, the stability of the Hungarian algorithm is much weaker than the GS algorithm. Therefore, the proposed GS algorithm with acceptable performance and computational complexity is suitable in this work.
\begin{figure}
  \centering
  \includegraphics[width=0.45\textwidth]{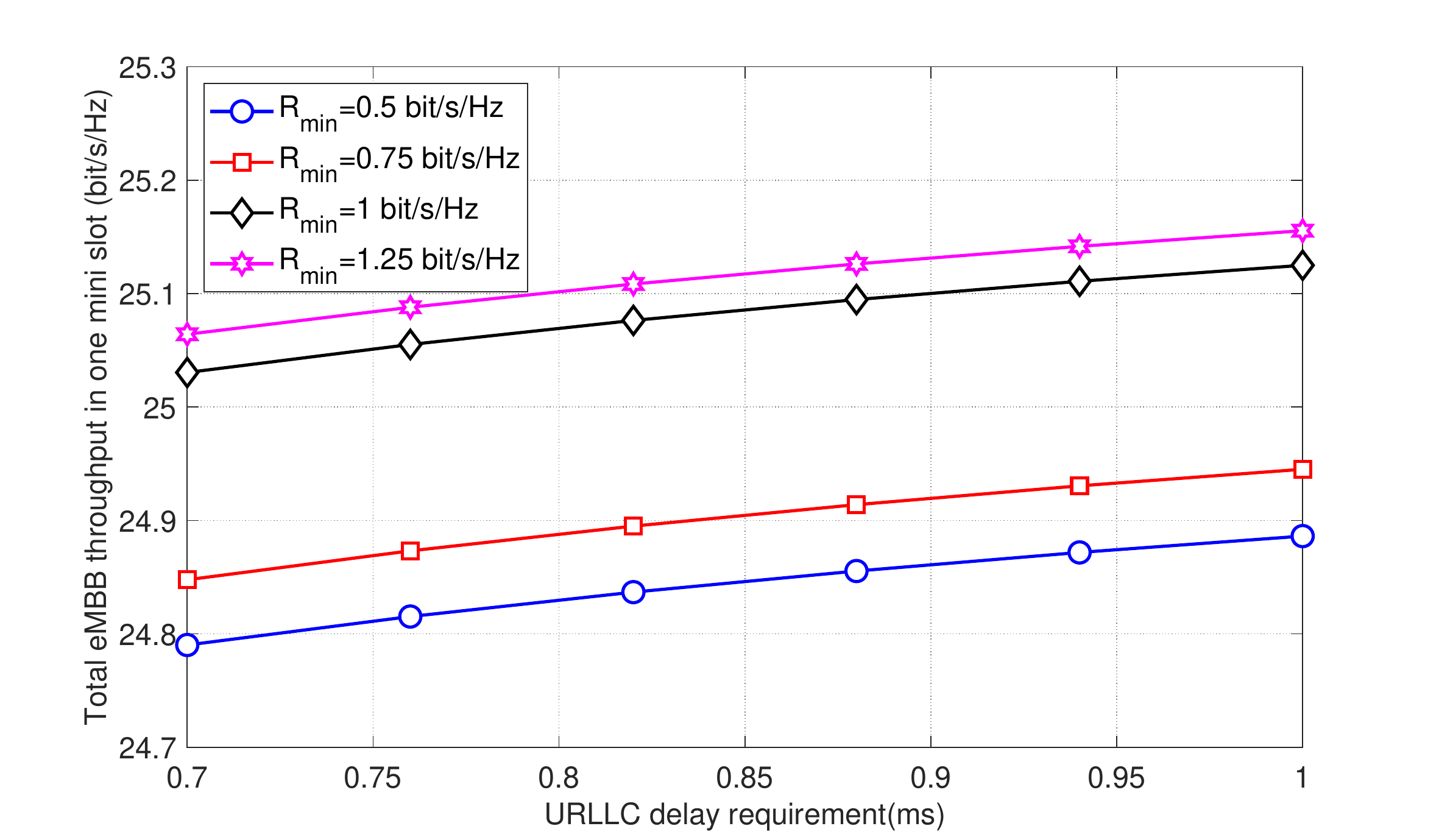}\\
  \caption{Performance comparison under different QoS requirements.}\label{fig:3}
\end{figure}

Fig. \ref{fig:3} shows the effect of the URLLC and eMBB users' QoS requirements on the eMBB throughput performance in the whole transmission period. In this simulation, we let two URLLC users arrive to the system. From this figure, we can find that the total eMBB throughput increases with the URLLC delay requirements.
This is because the more tolerant of the delay requirements, the lower power consumptions, and thus the BS can use the excessive power to improve the data rate of eMBB users. On the other hand, under fixed URLLC delay requirements, the performance of all eMBB users increase with the increment of eMBB users' data rate requirements.

\begin{figure}
  \centering
  \includegraphics[width=0.45\textwidth]{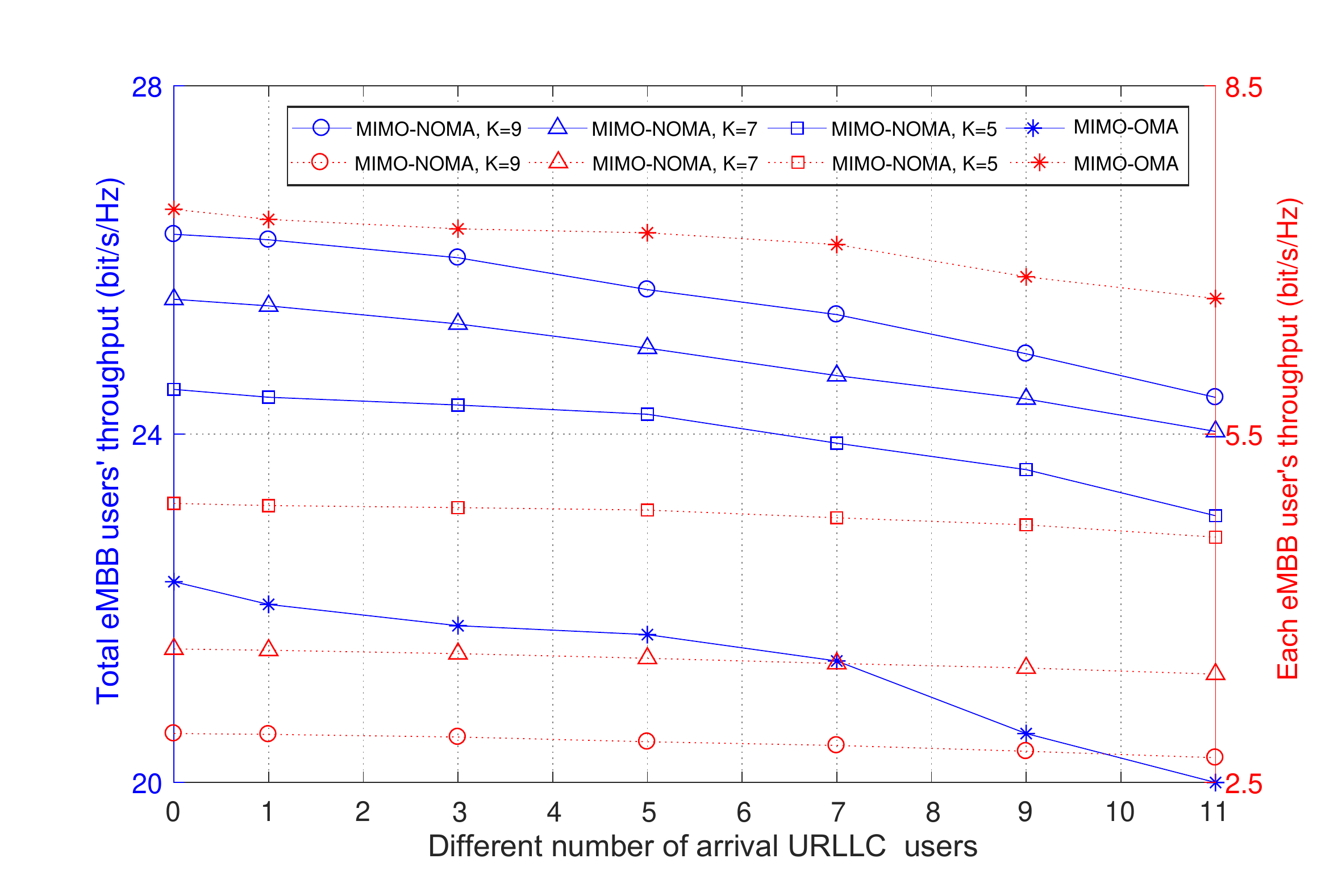}\\
  \caption{Performance comparison among different number of punctured URLLC users.}\label{fig:4}
\end{figure}

Fig. \ref{fig:4} presents the total eMBB users' throughput and each eMBB user's throughput in a transmission period under different number of arrival URLLC users. The MIMO-OMA scheme is selected as a benchmark.
From this figure, the MIMO-OMA scheme provides the highest per-user eMBB throughput, but the lowest total eMBB throughput.
This is because the MIMO-OMA scheme would connect fewer users than the MIMO-NOMA scheme to eliminate interference, and the MIMO-NOMA scheme has higher spectrum utilization.
Furthermore, the total eMBB performance decreases with the increased arrival URLLC users, since more URLLC users would preempt more eMBB resources. Under the fixed number of arrival URLLC users,
the total eMBB performance increases with $K$ due to the provided multiuser diversity gain.

\begin{figure}
  \centering
  \includegraphics[width=0.45\textwidth]{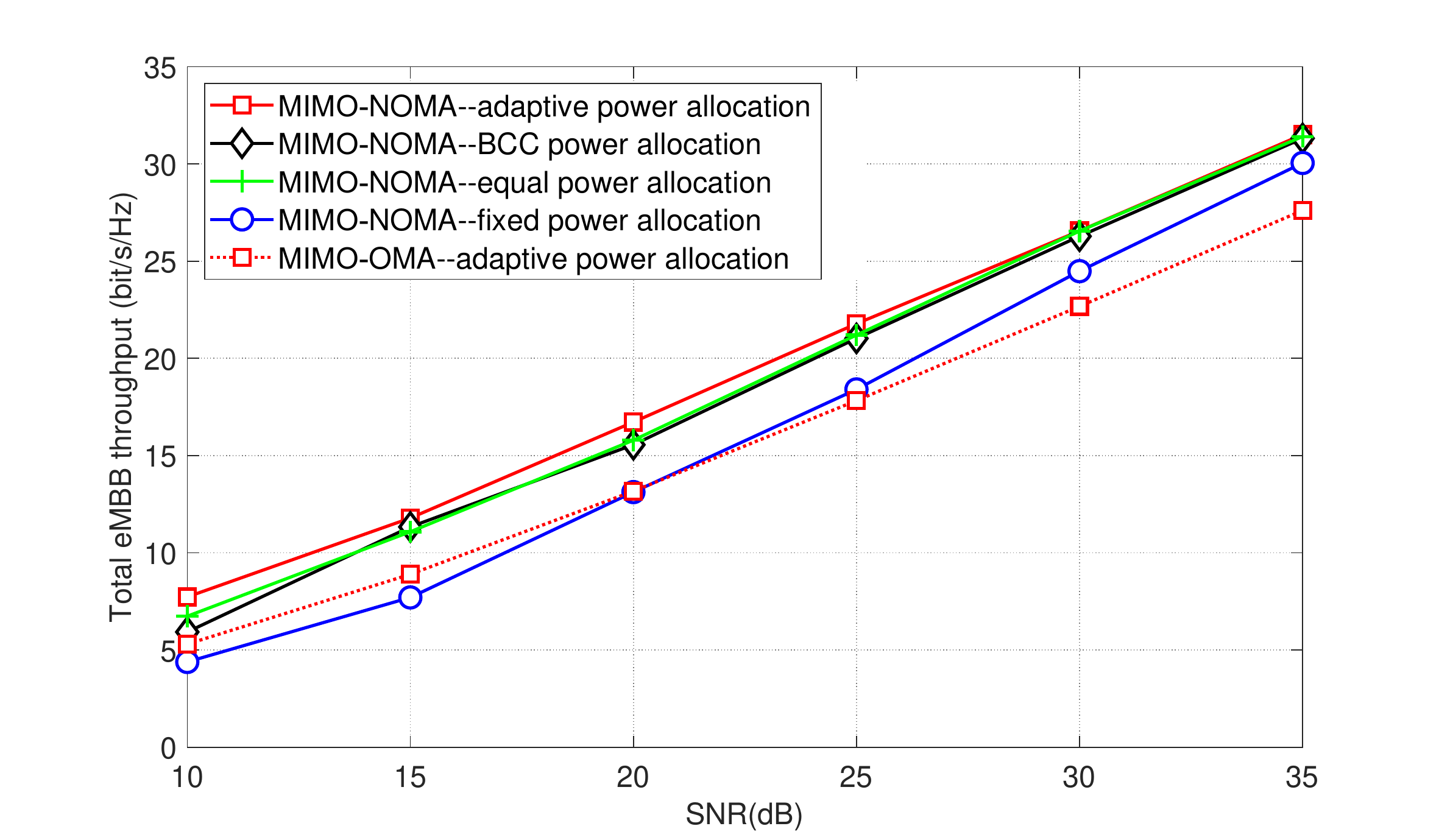}\\
  \caption{Performance comparison among different power allocation schemes.}\label{fig:5}
\end{figure}
\begin{figure}
  \centering
  \includegraphics[width=0.45\textwidth]{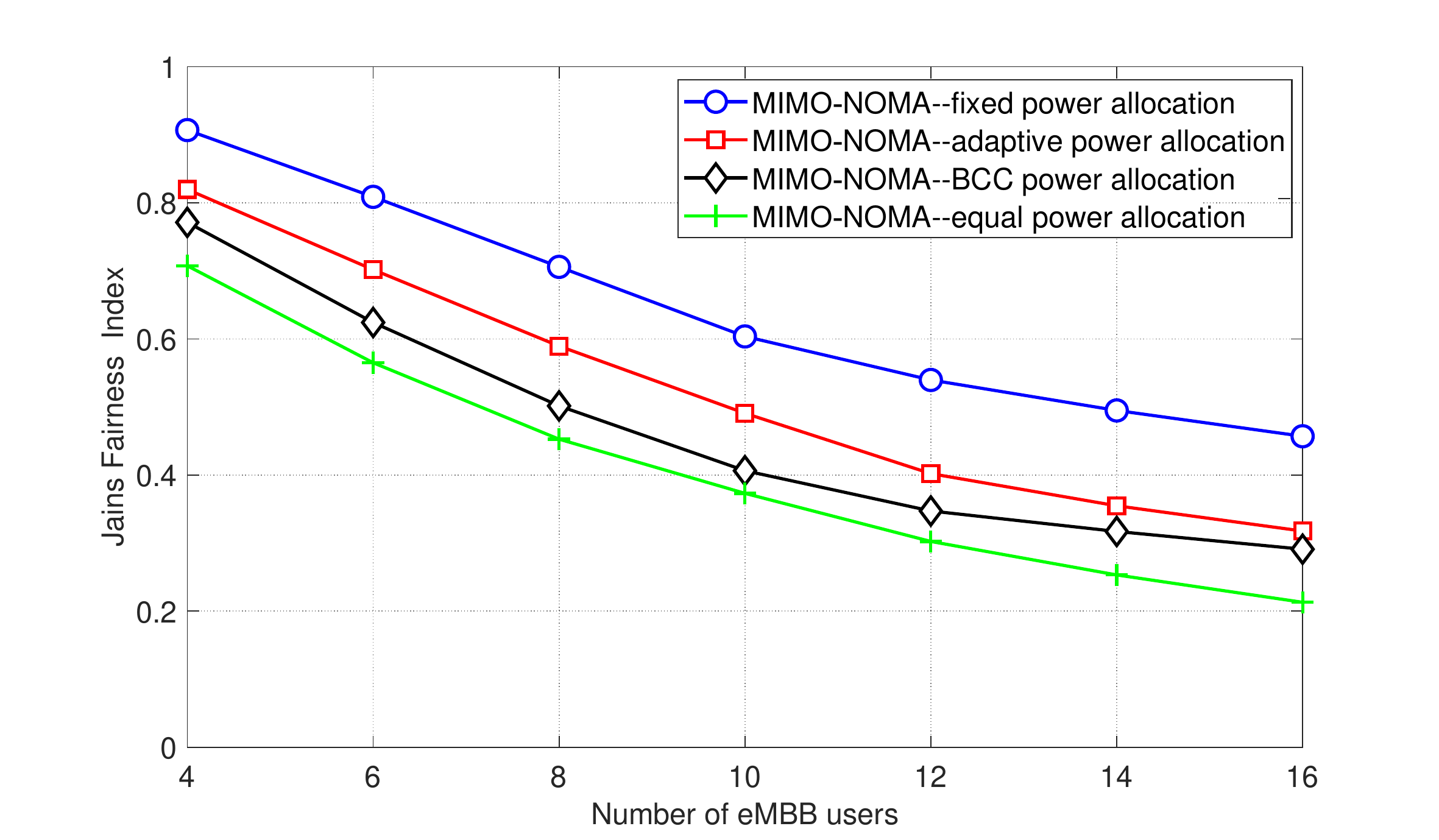}\\
  \caption{User fairness index under different number of eMBB users.}\label{fig:6}
\end{figure}

In Fig. \ref{fig:5}, we compare the proposed MIMO-NOMA based adaptive power allocation scheme with four baseline schemes, namely the equal power allocation scheme and the fixed power allocation scheme in \cite{fix_power}, the best channel condition (BCC) power allocation scheme in \cite{BCC_power}, and the MIMO-OMA based adaptive power allocation scheme in \cite{oma}. Note that the fixed power allocation scheme distributes ${\frac{k}{\sum_{k \in \mathcal{L}_{m}}{k}}}\frac{L_{m}}{K}$ to the $k$-th user in $m$-th cluster to due with the SIC.
In the BCC power allocation scheme, except for the user with the best channel condition, the achievable transmission rates for all the other users in the same cluster are set to satisfy the minimum QoS requirement. Here, we let $R_{m,k}^{\min}=0.2$ bit/s/Hz.
From this figure, we can find that the proposed adaptive power control scheme has the best performance. This is because the adaptive power control scheme can adjust the power allocation scheme based on the channel conditions and system parameters.  As the SNR increases, the equal power control scheme, the BCC power allocation scheme, and the proposed scheme almost have the same performance, however, the fixed allocation scheme has a clear performance decreasing. It is due to the equal power control scheme and the BCC power allocation scheme force the user with the largest channel gain to achieve the relatively high power compared to the proposed scheme, which can improve the spectral efficiency. However, both the equal power control scheme and the BCC power allocation scheme can not guarantee users' fairness well, which is shown in Fig. \ref{fig:6}. The fixed power allocation scheme has a worse performance than the equal power allocation scheme, since the fixed allocation scheme allocates more power to the user with a weak channel gain, resulting in a lower power utilization efficiency. Hence, it makes sense to select a proper non-orthogonal power optimization scheme according to the service characteristics in MIMO-NOMA systems. Moreover, it is also intuitively that MIMO-NOMA can significantly outperform MIMO-OMA under reasonably power allocation schemes.

Fig. \ref{fig:6} shows the user fairness under different number of eMBB users. We compare the proposed adaptive power allocation scheme with the equal power allocation scheme, the fixed power allocation scheme, and the BCC power allocation scheme. According to \cite{userclustering_fairness}\cite{bookjainfair}, Jain's fairness index is obtained by
$\left(\sum_{m\in \mathcal{M}}
\sum _{k \in \mathcal{L}_m}R_{m,k}^{(i)}\right)^2/{\left|{\mathcal{S}^{(i)}_{~\!e}}\right |}\left(\sum_{m\in \mathcal{M}}
\sum _{k \in \mathcal{L}_m}{R_{m,k}^{(i)~2}}\right)$,
 where $R_{m,k}^{(i)}$ is the achievable data rate of eMBB user $k$ in the $m$-th cluster at mini slot $i$.
 With the increment of eMBB users, the fairness index among them decreases due to the competition. The fixed power allocation scheme achieves higher user fairness than the proposed adaptive power allocation, since the proposed adaptive power allocation improve the spectrum utilization at the cost of fairness. By combining Fig. \ref{fig:5} and Fig. \ref{fig:6}, we can conclude that the proposed adaptive power allocation scheme can significantly improve the performance at an acceptable fairness level.

\begin{figure}
  \centering
  \includegraphics[width=0.45\textwidth]{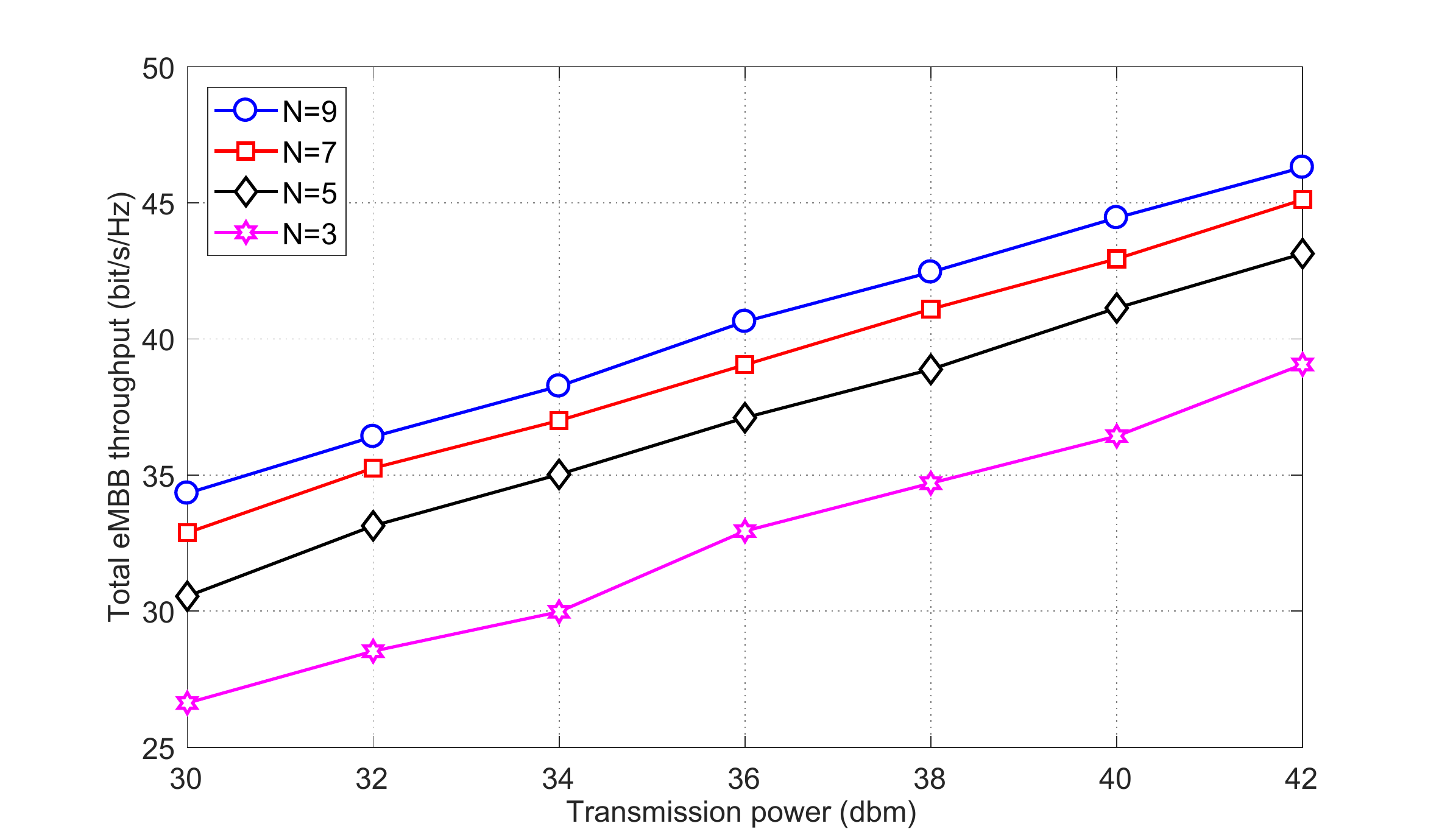}\\
  \caption{Performance comparison under different number of antennas and transmission power. }\label{fig:7}
\end{figure}
Fig. \ref{fig:7} shows the system data rate under different transmission power and user antennas.
From this figure, we can observe that the total eMBB throughput in a transmission period of our proposed system is improved with the increment of transmit power. Moreover, the total data rate increases with the increment of user antennas, since the dimension of the null space $\mathbf{U}_{l,k}^{(i)}$, defined in \eqref{9}, increases and the ability of detection vector to eliminate interference improves.

\begin{figure}
  \centering
  \includegraphics[width=0.45\textwidth]{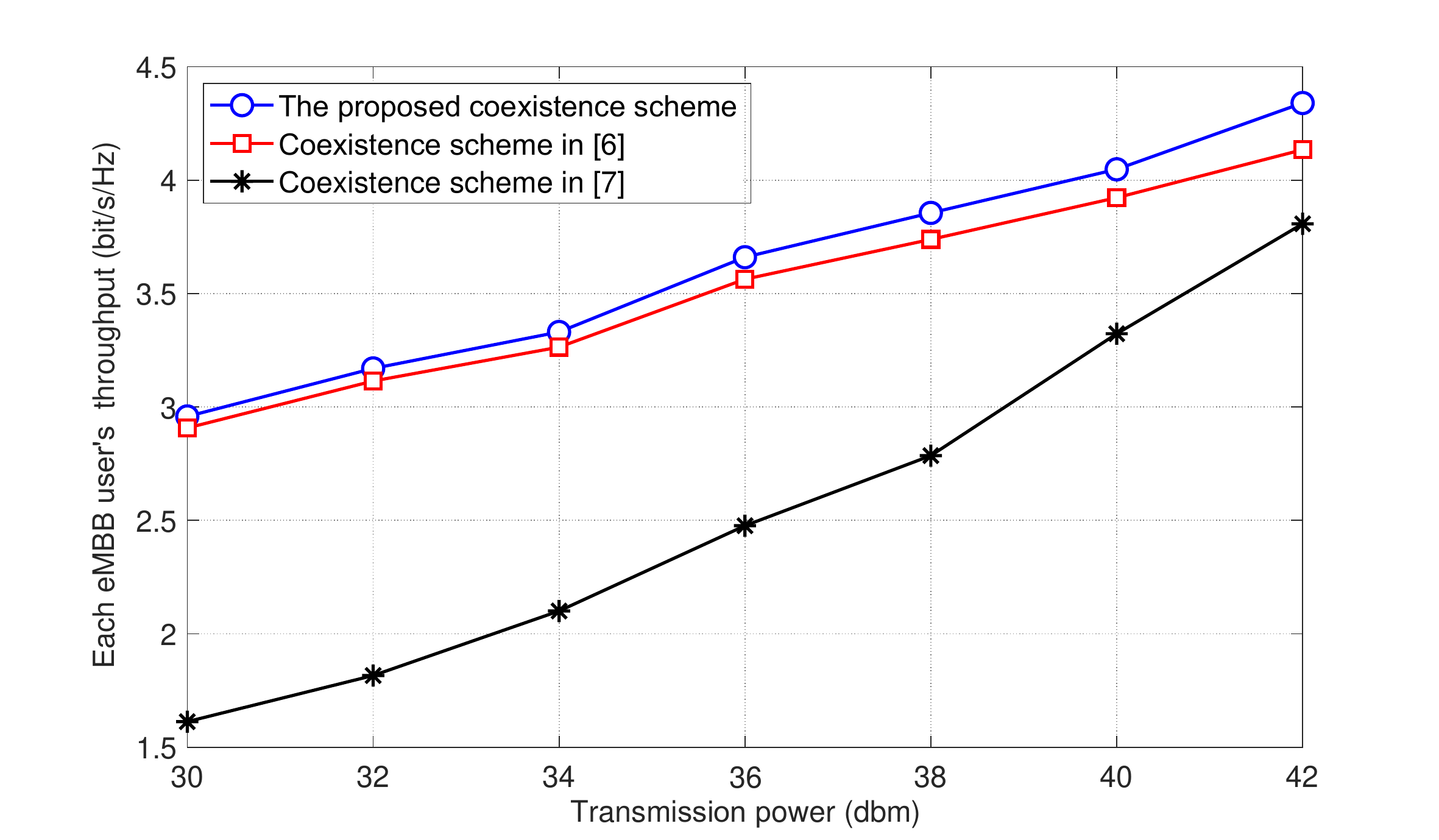}\\
  \caption{Performance comparison among different URLLC and eMBB coexistence mechanisms.}\label{fig:8}
\end{figure}

Fig. \ref{fig:8} illustrates the performance comparison among different URLLC and eMBB coexistence mechanisms.
The linear model for eMBB throughput loss under superposition mechanism in \cite{urllc-embb-anand} (LM-SM) and
the OMA based spatial coexistence mechanism in \cite{chenlaoshi} are selected as two benchmarks.
In the LM-SM scheme, we adopt the superposition mechanism for URLLC scheduling and leverage the GS method proposed in our paper to perform the user selection in the clustered MIMO-NOMA system. Moreover, the rate loss of the superimposed eMBB users is calculated through the linear model proposed in \cite{urllc-embb-anand}.
In the coexistence scheme in \cite{chenlaoshi}, each eMBB user occupies one spatial resource. Any arriving URLLC user can select an eMBB user to replace and occupy the entire spatial dimension under the puncturing mechanism.
From this figure,the proposed coexistence scheme performs best. Moreover, the LM-SM scheme has better performance than the coexistence scheme in \cite{chenlaoshi}. The reason is that the resource utilization in \cite{chenlaoshi} is insufficient. Moreover, QoS requirements of both URLLC and eMBB users are ignored in the user scheduling program in \cite{chenlaoshi}.

\section{Conclusion}\label{E}
This paper proposes a cluster-based MIMO-NOMA technology to the URLLC and eMBB coexistence network.
Before transmission, we propose a dynamic eMBB user clustering mechanism in MIMO-NOMA systems to strike a balance between the system performance and the computational complexity. Thereafter, we adopt a punctured scheduling scheme, which allows arriving URLLC users interrupt parts of associated eMBB users. An eligible and optimal interrupted eMBB users can be found through a GS matching algorithm with low computational complexity. For further spectrum utility improvement, we propose an iterative power allocation algorithm based on the SCA and D.C. programming. Our numerical results demonstrate a notable performance improvement of the proposed mechanism over the baseline methods in the presented indicators.

\appendices
\section{Convergence proof of Algorithm \ref{power}}\label{proofdc}
We first introduce the following stationary point definition according to \cite{conv}.
\begin{definition}
Let $f: \mathcal{D}\rightarrow \mathbb{R}$ be a function where $\mathcal{D}\subseteq \mathbb{R}^m$ is a convex set. The point $x$ is a stationary point of $f(\cdot)$ if $f'(x;d)\geq 0$ for all $d$ such that $x+d \in \mathcal{D}$. Here, $d$ is the distance of a point from a set.
\end{definition}

Thereafter, the conditions to setup the approximation functions to locally approximate the non-convex constraints can be summarized as the following assumption \cite{jianhua-tang-urllc-embb-slice}.
\begin{assumption}
A function $u(x,y)$ is called as the  \emph{approximation function} for the non-convex function $f(x)$, when the following conditions satisfy:
\begin{equation}\label{A1}
u(x,x)=f(x), \forall x,
\end{equation}
\begin{equation}\label{A2}
u(x,y)\geq f(x), \forall x, y,
\end{equation}
\begin{equation}\label{A3}
\frac{\partial u(x,y)}{\partial x}|_{x=y}=\triangledown f(x)|_{x=y}, \forall x,
\end{equation}
\begin{equation}\label{A4}
u(x,y)~\mathrm{is~continuous~in}~(x, y),
\end{equation}
\begin{equation}
u(x,y)~\mathrm{is~convex~in}~x.
\end{equation}
\end{assumption}

Define the non-convex function of $\mathcal{P}_4$ as $f(\bm{\gamma}^{(i)})={G_{1}\left ({{\bm{\gamma}^{(i)}} }\right) - {G_{2}}\left ({{\bm{\gamma}^{(i)}} }\right)}$ and the approximation function of $\mathcal{P}_5$ as $u(\bm{\gamma}^{(i)},\bm{\gamma}^{(i),p})=G_{1}\!\left ({{\bm{\gamma}^{(i)}} }\right) \!-\! {G_{2}}\!\left ({{\bm{\gamma}^{(i),p}} }\right)\!-\!{\nabla _{\bm{\gamma}^{(i)}}}{G_{2}}\!{\left ({{\bm{\gamma}^{(i),p}} }\right)^{\mathrm {T}}}\left ({{\bm{\gamma}^{(i)}} \!- \! {\bm{\gamma}^{(i),p}} }\right)$.

We can observe the following series of relationships as
\begin{multline}\label{budengshi}
f({\bm{\gamma}^{(i),p+1}}) \overset{(a)}{\leq} u(\bm{\gamma}^{(i),p+1}, \bm{\gamma}^{(i),p}) \\
\overset{(b)}{\leq} u({\bm{\gamma}^{(i),p}}, \bm{\gamma}^{(i),p}) \overset{(c)}{=} f({\bm{\gamma}^{(i),p}}),
\end{multline}
where $(a)$ is due to \eqref{A2}, $(b)$ follows from the optimality of
$\bm{\gamma}^{(i),p+1}=\arg\underset{\bm{\gamma}^{(i)}}{\min}~ u({\bm{\gamma}^{(i)}},{\bm{\gamma}^{(i),p}})$,
and $(c)$ is due to \eqref{A1}.

A straightforward consequence of \eqref{budengshi} is that the sequence of $P_4$ values are non-increasing at each iteration, which can be formulated as
\begin{equation}\label{nonincreasing}
f({\bm{\gamma}^{(i),0}})\geq f({\bm{\gamma}^{(i),1}}) \geq f({\bm{\gamma}^{(i),2}}) \geq \cdots.
\end{equation}
Assume that there exists a subsequence $\left\{\bm{\gamma}^{(i),p_j}\right\}$ converging to a limit point $z$. Combing \eqref{A1}, \eqref{A2}, and \eqref{nonincreasing}, we can derive that
\begin{multline}\label{R9}
u({\bm{\gamma}^{(i),p_{j+1}}},{\bm{\gamma}^{(i),p_{j+1}}})=f({\bm{\gamma}^{(i),p_{j+1}}})\leq f({\bm{\gamma}^{(i),p_{j}+1}}) \\
\leq u({\bm{\gamma}^{(i),p_{j}+1}},{\bm{\gamma}^{(i),p_{j}}})\leq
u({\bm{\gamma}^{(i)}},{\bm{\gamma}^{(i),p_{j}}}).
\end{multline}

Letting $j\rightarrow \infty$, we can obtain from \eqref{R9} that
\begin{equation}
u(z,z)\leq u({\bm{\gamma}^{(i)}}, z),
\end{equation}
which implies that
\begin{equation}
u'({\bm{\gamma}^{(i)}}, z; d)|_{{\bm{\gamma}^{(i)}}=z}\geq 0,~ \forall d\in \mathbb{R}^m.
\end{equation}

With the assumption of \eqref{A3}, we have
\begin{equation}
f'(z; d)\geq 0,~ \forall d\in \mathbb{R}^m,
\end{equation}
which implies that $z$ is a stationary point of $f(\cdot)$.
It ends the proof.

\ifCLASSOPTIONcaptionsoff
  \newpage
\fi

%\bibliographystyle{IEEEtran}
%\bibliography{Doc5}

%\end{IEEEbiography}


%\newpage


%\vfill


\end{document}